\documentclass[a4paper,fleqn,usenatbib,useAMS]{mnras}

    \usepackage{float}
    \usepackage{graphicx}    
    \usepackage{amsmath}   
    \usepackage{amssymb}    
    \usepackage{bm}      
    \usepackage{subfloat}
    \usepackage[T1]{fontenc}
    \usepackage{ae,aecompl}
    \usepackage{newtxtext,newtxmath}
	\usepackage{color}

    \newcommand{\nustar}{\emph{NuSTAR}}
   \newcommand{\ngc}{NGC\ 3718}
		\title[X-ray reflection in the LLAGN NGC\,3718]{Constraining X-ray reflection in the low-luminosity AGN NGC\,3718 using \emph{NuSTAR} and \emph{XMM--Newton}}
	    \author[Y. Diaz et al.]{
Y. Diaz,$^{1}$\thanks{E-mail: yaherlyn.diaz@postgrado.uv.cl}, P. Ar\'evalo, $^{1}$ L. Hern\'andez-Garc\'ia,$^{1}$ L. Bassani,$^{2}$ A. Malizia,$^{2}$ O. Gonz\'alez-\newauthor Mart\'in,$^{3}$ C. Ricci,$^{4,5,6}$ G. Matt,$^{7}$ D. Stern,$^{8}$ D. May, $^{9}$ A. Zezas, $^{10}$ F. E. Bauer. $^{11,12,13}$
    \\
    	$^{1}$ Instituto de F\'isica y Astronom\'ia, Facultad de Ciencias,Universidad de Valpara\'iso, Gran Bretana No. 1111, Playa Ancha, Valpara\'iso, Chile 
		\\
		$^{2}$ OAS-INAF, Via P. Gobetti 101, 40129 Bologna, Italy \\
		$^{3}$ Instituto de Radioastronom\'ia and Astrof\'isica (IRyA-UNAM), 3-72 (Xangari), 8701, Morelia, Mexico \\
		$^{4}$ N\'ucleo de Astronom\'ia de la Facultad de Ingenier\'ia, Universidad Diego Portales, Av. Ej\'ercito Libertador 441, Santiago, Chile \\
		$^{5}$ Kavli Institute for Astronomy and Astrophysics, Peking University, Beijing 100871, China \\
		$^{6}$ George Mason University, Department of Physics \& Astronomy, MS 3F3, 4400 University Drive, Fairfax, VA 22030, USA\\
		$^{7}$ Dipartimento di Matematica e Fisica, Universit\'a degli Studi Roma Tre, via della Vasca Navale 84, 00146 Roma, Italy \\
		$^{8}$ Jet Propulsion Laboratory, California Institute of Technology, Pasadena, CA 91109, USA \\
		$^{9}$ Instituto de Astronomia, Geof\'isica e Ciencias Atmosf\'ericas, Universidade de Sao Paulo, 05508-090, Sao Paulo, SP, Brazil \\
		$^{10}$ University of Crete, Department of Physics, Voutes University Campus, GR-71003 Heraklion, Greece \\
		$^{11}$ Instituto de Astrof{\'{\i}}sica and Centro de Astroingenier{\'{\i}}a, Facultad de F{\'{i}}sica, Pontificia Universidad Cat{\'{o}}lica de Chile, Casilla 306, Santiago, Chile \\
		$^{12}$ Millennium Institute of Astrophysics (MAS), Nuncio Monse{\~{n}}or S{\'{o}}tero Sanz 100, Providencia, Santiago, Chile \\
		$^{13}$ Space Science Institute, 4750 Walnut Street, Suite 205, Boulder, Colorado 80301
		}
	
\begin{document}

    \date{Accepted 2020 June 12. Received 2020 June 2; in original form 2020 February 7}

    \pagerange{\pageref{firstpage}--\pageref{lastpage}} \pubyear{2020}

    \maketitle

    \label{firstpage}

    \begin{abstract}
    
One distinctive feature of low-luminosity active galactic nuclei (LLAGN) is the relatively weak reflection features they may display in the X-ray spectrum, which can result from the disappearance of the torus with decreasing accretion rates. Some material, however, must surround the active nucleus, i.e., the accretion flow itself and, possibly, a flattened-out or thinned torus. In this work, we study whether reflection is indeed absent or undetectable due to its intrinsically weak features together with the low statistics inherent to LLAGN. Here we focus on  NGC\,3718 ($L/L_{\rm Edd}\sim10^{-5}$) combining observations from \emph{XMM--Newton} and the deepest to date \emph{NuSTAR} (0.5--79 keV) spectrum of a LLAGN, to constrain potential reflectors, and analyze how the fitted coronal parameters depend on the reflection model.
We test models representing both an accretion disc (Relxill) and a torus-like (MYTorus and Borus) neutral reflector. From a statistical point of view, reflection is not required, but its inclusion allows to place strong constraints on the geometry and physical features of the surroundings: both neutral reflectors (torus) tested should be Compton thin ($N_H<10^{23.2}$cm$^{-2}$) and preferentially cover a large fraction of the sky. If the reflected light instead arises from an ionized reflector, a highly ionized case is preferred. These models produce an intrinsic power-law spectral index in the range [1.81--1.87], where the torus models result in steeper slopes. The cut-off energy of the power-law emission also changes with the inclusion of reflection models, resulting in constrained values for the disc reflectors and unconstrained values for torus reflectors.

    \end{abstract}

    \begin{keywords}
    galaxies -- individual: NGC\,3718, galaxies -- nuclei, galaxies -- X rays: galaxies -- variability
    \end{keywords}

    \section{Introduction}
    \label{sec:intro}
    
    Active galactic nuclei (AGN) emit over the entire electromagnetic spectrum and are powered by accretion onto a supermassive black hole  \citep{1984Rees}. They are divided into two classes depending on the presence (type 1) or absence (type 2) of broad emission lines observed superimposed with narrow emission lines in their optical spectra. According to the unification model (UM), the different classes of observed AGN are related to the existence of dust and gas toroid-like structure, popularly called the torus, surrounding the central engine \citep{1993Antonucci}. If the AGN is seen face on, there is a direct view to the accretion disc and the broad line region (BLR)
    , giving rise to a type 1 object. If seen edge-on, the torus obscures the UV and optical light from the accretion disc and the BLR, being classified as a type 2 object.   
 
 Orientation is not the sole explanation for the different types of AGN. Differences in accretion rate are also important, with Narrow-Line Seyfert 1 galaxies (NLSy1s) at one extreme of this parameter and at least some low-ionization nuclear emission-line region (LINERs) galaxies at the other. It has become apparent that the physical extent of the obscuring material (i.e., the torus) is itself a function of the accretion rate \citep[e.g.,][]{2017Riccii}. These authors found that the fraction of obscured sources, or covering factor of the Compton-thin circumnuclear material, grows with the accretion rate and then exhibits a sharp decline at an Eddington ratio ($R_{\rm Edd}$=$L_{\rm Bol}/L_{\rm Edd}$, where $L_{\rm Bol}$ is the bolometric luminosity and $L_{\rm Edd}$ is the Eddington-limit luminosity) of $\sim$10$^{-2}$. This value of $R_{\rm Edd}$ corresponds to the Eddington limit for dust. According to this result, the probability of an AGN to be obscured is mostly driven by the Eddington ratio, resulting in a radiation-regulated unification model. The decline of covering fractions at high accretion rates can be explained by the AGN being powerful enough to radiatively blow away the circumnuclear material, while the decline at low accretion rates points either to a lower ability to inflate a torus (e.g., \citealt{2008Elitzur}) or to smaller amounts of circumnuclear gas leading to lower accretion rates. Current open questions include the nature of the torus, and its dependence on luminosity, black hole mass and galaxy evolution \citep[e.g.,][]{2015Netzer}.  
    
     The primary X-ray emission originates in a corona close to the accretion disc \citep[e.g.,][]{1993haardt} and is well represented by a power-law model \citep[e.g.,][]{Nandra1994, Nandra1997, Risaliti2002, cappi2006, 2009Gonzalez}. When this X-ray continuum is scattered by the surrounding gas, new features are imprinted in the spectrum, producing  fluorescent emission lines, most notably Fe K$\alpha$ 6.4 keV, and a broad hump-like continuum peaking around 10--30 keV \citep[e.g.,][]{piro1990}. The relative strength of these two features is related to the column density of the scattering gas, while their overall flux is proportional to the gas covering fraction as seen from the central engine. Therefore, a careful analysis of the scattered X-ray spectrum can reveal the presence and properties of the obscuring torus even in unobscured AGN. Therefore, the X-ray spectrum is an useful tool to study the properties of obscuration in AGN because it gives information on the circumnuclear material even if it does not lie in the line of sight to the corona.
    
    Studying the properties of the reflector spectrum is a difficult task because it is affected by additional absorption, the uncertain spectral slope of the power-law and a possible high energy cut-off ($E_{\rm cut}$), which mimics the curvature of the scattered light, and by the additional  contribution of ionized reflection (e.g., from the accretion disc). For this reason, hard X-ray observations of AGN, above 10 keV, are of paramount importance to constrain the complete reflection spectrum. A careful analysis of the reflection component is also crucial to study the accretion mechanism. Given the partial degeneracy between the curvature of the scattered spectral component and the curvature of the coronal emission, without a good constraint on reflection it is difficult to estimate the real shape of the coronal spectrum. In particular, low-luminosity AGN (LLAGN) are thought to have a different accretion mechanism compared to more powerful AGN \citep{yamaoka2005, yuan2007, 2009Gu, 2011younes, 2011xu, 2016Lore, 2018She}. 
    The standard accretion disc model \citep{1973shakura,  1999Pkora} can successfully explain the AGN power in the regime of high accretion rates ($R_{\rm Edd}$ > 10$^{-3}$). The standard disc is geometrically thin and optically thick, with viscous dissipation balancing radiative cooling locally.  On the other hand, in the low-accretion regime ($R_{\rm Edd}$ $\leq$ 10$^{-3}$ ), the standard cool disc model is no longer able to fit the observations. These low-luminosity AGN are found to be radiatively inefficient \citep{2008ho}. In this regime advection-dominated accretion flows (ADAFs; e.g, \citealt{1994narayan}) are expected.

LINERs are LLAGN ($L_{\rm Bol} \sim $ 10$^{\rm 39}$-10$^{\rm 42}$ erg s$^{\rm -1}$, \citealt{2008ho}) and are characterized by strong low ionization optical emission lines such as [N \,{\sc II}] $\lambda$6584$\AA$, [O \,{\sc I}] $\lambda$6300$\AA$, and [S\,{\sc II}] $\lambda$6731$\AA$ being relatively stronger than higher ionization emission lines \citep{1980Heckman}. In these objects the ionization mechanism is still not clear \citep{2008ho}. 
    The spectral energy distribution (SED) does not show the big blue bump of more luminous Seyferts and quasars, a classical signature of the innermost, geometrically thin accretion disc. However, a truncated thin disc is necessary to explain the big red bump observed---the prominent mid-IR peak and the steep fall-off of the spectrum in the optical-UV region, which conforms to the scenario where the innermost accretion disc has been replaced by an ADAF. \citep{2008ho}). 
       
       One of the distinctive features of LLAGN is the low level of reflection features that they display in their X-ray spectra \citep[e.g.,][]{2019younes, 2019natalia}, together with the small contribution from torus emission in the mid-IR band \citep[e.g.,][]{2017Gonzalez-martin}. This small amplitude complicates the study of reflection in LLAGN, requiring high quality, high energy data. These observations can be explained in the context of a disappearing torus with decreasing accretion rates \citep{2008Elitzur}. Some material, however, must surround the X-ray emitting region, e.g., the rest of the accretion flow itself and possibly a flattened-out or thinned remnant of the torus and the reflection off these structures should be visible in a sufficiently precise X-ray spectrum.  The goal of our work is to constrain the properties of these remnant structures to establish whether the observed low reflection implies a completely clear sky as viewed from the corona, or if a large torus or untruncated disc are still compatible with the observations.

    In this work, we use simultaneous observations by \emph{XMM-Newton} and \emph{NuSTAR} to study in detail the LLAGN NGC\,3718, a nearby galaxy at redshift\footnote{https://ned.ipac.caltech.edu/} $z$=0.003. This galaxy is classified as a type 1.9 LINER (optical classification, e.g., \citealt{1997Ho, 2018cazzoli}), with a black-hole mass $\log(M_{BH})=7.85\pm$1.42 given by \citealt{2014Lore} and determined using the correlation between stellar velocity dispersion and black-hole mass of \citet{tremaine2002}, where the stellar velocity dispersion used was $\sigma=169.9$ km s$^{-1}$ \citep{2009bho}. It has a low accretion rate $R_{\rm Edd} \sim  4\times 10^{-5}$, estimated from its mass, its 2--10 keV X-ray luminosity (log [$L_{X}$/(erg s$^{-1}$)]=40.4, \citealt{Satyapal2005}) and bolometric correction factor ($L_{\rm Bol}/L_{\rm 2.0-10.0 keV} =15$) appropriate for low-luminosity AGN \citep{lusso2012}. 
    
    The high-quality X-ray spectra allow us to analyze how the fitted coronal parameters (photon index $\Gamma$ and $E_{\rm cut}$) depend on the reflection model, as well as constrain the torus/reflector properties in this low accretion rate regime. This paper is organized as follows: in Sect. \ref{sec:2} we present details of the observations and data analysis procedures. The variability and spectral results are reported in Sect. \ref{sec:3}. The implications of our X-ray spectral analysis are discussed in Sect. \ref{sec:4}. Finally, a summary of our findings is presented in Sect. \ref{sec:5}.

    \section{Observations and data reduction}
    \label{sec:2}

    The \emph{XMM--Newton} observations were performed on 2017 October 24 using the medium filter, with Full Frame mode in the EPIC-PN \citep{2001turner} and Large Window mode in both EPIC-MOS \citep{2001struder} cameras. The \emph{XMM--Newton} data were processed with SAS version 16.1.0, using the metatasks {\sc epproc} and {\sc emproc} and events were selected with the task {\sc evselect}. The spectra were constructed from cleaned events files where the flaring times were removed by applying a threshold of 0.7 counts per second on the PN 10.0-12.0 keV count rate integrated over the entire field of view (FOV). Of the original live time of 28.3 ks in the PN CCD 4, only 18.2 ks livetime were used for the spectral analysis.  Source events were selected for each detector from circular regions of 40 arcsec in radius, centered on the target. Background events were selected from source-free regions of equal area on the same chip as the source, approximately 100 arcsec away.  The same good time intervals were applied to PN, MOS1 and MOS2 data, resulting in source+background counts of 5440 for the PN, 1700 for MOS1 and 2100 for MOS2; the estimated  source fractions were 91\%, 92.5\% and 96\%, respectively.  Spectral channels were grouped with the SAS task {\sc specgroup} to contain a minimum of 25 counts per bin. Spectral response (RMF) and ancillary (ARF) files were created using the tasks {\sc rmfgen} and {\sc arfgen}.
    
    In case of the \emph{Nuclear Spectroscopic Telescope Array} (\emph{NuSTAR}, \citealt{2013harrison}), its two focal plane modules (FPMA and FPMB) operate in the energy range 3--79 keV. The observation was split into four segments spread within 10 days between 2017 October 24 and 2017 November 03, totalling almost 230 ks of exposure time. The \emph{NuSTAR} data were processed using {\sc nustardas}  v1.6.0, available in the \emph{NuSTAR} Data Analysis Software. The  event data files were calibrated with the {\sc nupipeline} task using  the  response files  from  the  Calibration  Database {\sc caldb} v.20180409  and  HEASOFT version 6.25.  With the {\sc nuproducts} script we generated both the source and background spectra, plus the ARF and RMF files. For  both  focal plane modules, we used a circular extraction region of radius 50 arcsec centered on the position of the source with a source-free. The background selection was made taking a region free of sources of twice the radius of the target.  Spectral channels were grouped with the {\sc ftools} task {\sc grppha} to have a minimum of 50 counts per spectral bin. The source is significantly detected in the 3--70 keV energy range. Details on the observations can be found in Table \ref{tab:tabla_obs1}.

   We also retrieved a high-energy spectrum for NGC 3718 from the \emph{Swift}/BAT 70 month All-sky Hard X-Ray Survey reported in \citet{2013baumgartner}, together with the corresponding response matrix. The data reduction and analysis for the \emph{Swift}/BAT 70 month All-sky Hard X-Ray Survey are based on the procedures used in the \emph{Swift}/BAT 22 All-Sky Hard X-ray Survey.   The complete  analysis  pipeline  is  described  in  the \emph{Swift}/BAT 22 All-sky Hard X-Ray Survey \citep{tueller2010}.
   This spectrum contains eight energy channels in the 14--200 keV energy range.

    \begin{table}
    
    \begin{center}
    
    \caption[Observational details]
    {Observational details. }
    \label{tab:tabla_obs1}
    \begin{tabular}{cccc}
    \hline \hline
    Telescope & Obs ID & Date & Exp. time \\
    \hline 
    \textit{NuSTAR} & 60301031002 & 2017/10/24	&  24.52/24.47  \\
     & 60301031004 & 2017/10/27	&  90.39/90.14 \\
     & 60301031006 & 2017/10/30	& 57.37/57.26   \\
     & 60301031008 &  2017/11/03  & 57.01/56.83  \\
     \hline
    \textit{XMM-Newton} & 0795730101 & 2017/10/24 & 18.2    \\
         \hline

    \textit{Swif}/BAT & - & 2004/12-2010/08 & 12.7$\times$10$^3$    \\

    \hline
    \end{tabular}
    \begin{minipage}{8cm}
    \textbf{Notes:} Instrument (Col. 1), obs ID (Col. 2), date (Col. 3), Net exposure times (Col.4)  
    represents the live time (in ks), of FPMA/FPMB for \emph{NuSTAR} and PN for \emph{XMM-Newton}. \emph{Swift/}BAT data correspond to the 70 month All-sky Hard X-Ray Survey reported in \citet{2013baumgartner}.
    \end{minipage}
    \end{center}
    \end{table}
	
   \section{Analysis and results}
    \label{sec:3}
    
    The spectral fitting process has two steps: \textbf{(1)} variability analysis of each \nustar\ exposure to study the possibility of combining them to increase the sensitivity and \textbf{(2)} modelling of the resulting spectra. 
    
	\subsection{Variability}
   
    We ran the task {\sc nuproducts} to construct light curves from the \nustar\ data in the energy range 3--70 keV with 500 s bin. 
    We subtracted the background light curves from the corresponding source light curves. The average count rates per segment, i.e., averaging FPMA and FPMB, range from $0.06 \pm{0.01}$ to $0.04\pm 0.01$, within $2\sigma$ of each other. The data are therefore consistent with no variations on a timescale of 10 days. 
    
\begin{figure}
    \resizebox{\hsize}{!}{\includegraphics{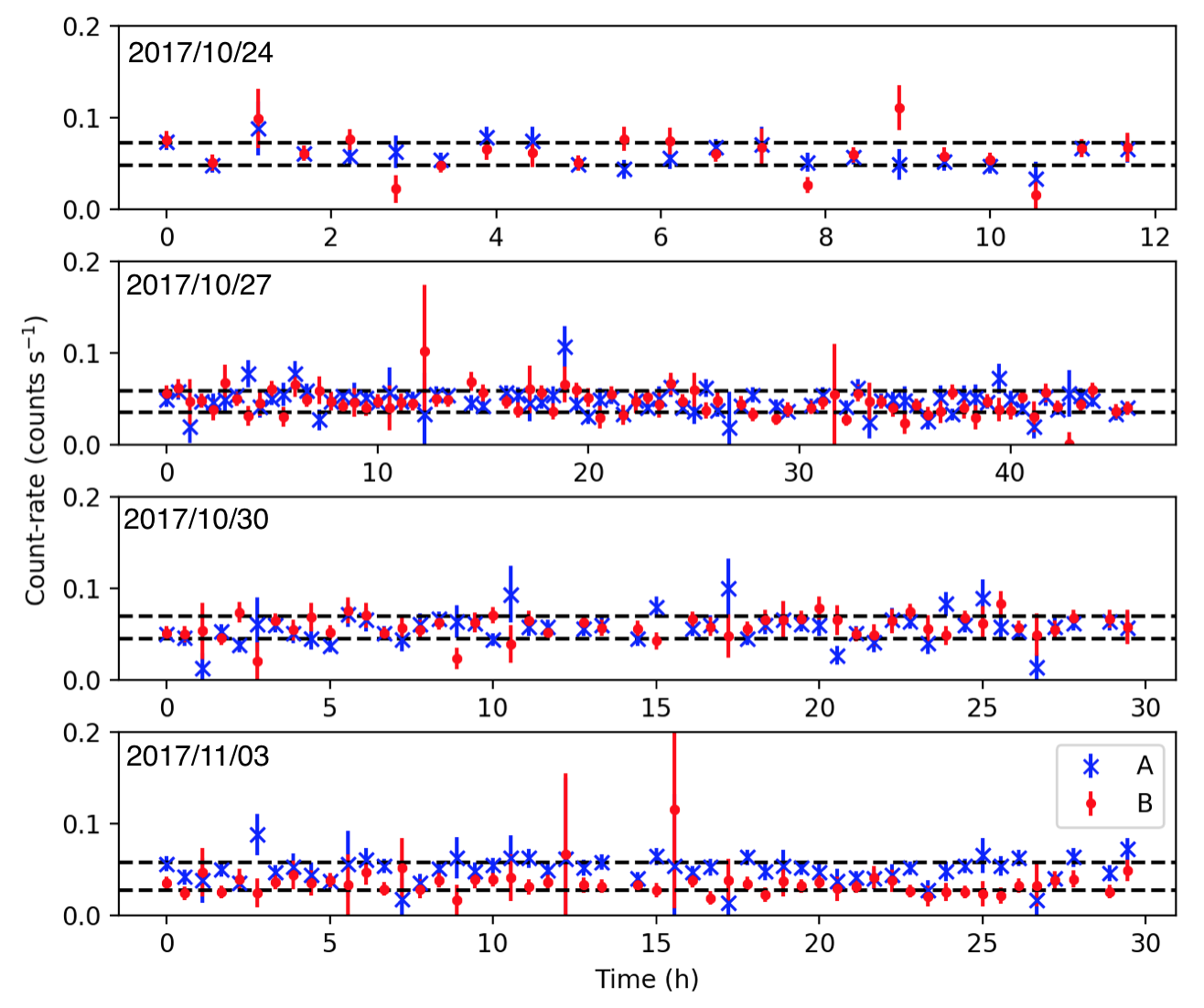}}
    \caption{Light curve of \emph{NuSTAR} data for NGC\,3718 with 500 s bin. Blue stars represent FPMA and red dots FPMB. The black dashed lines represent the 1$\sigma$ level from the average. The observations are separated by three days.}
     \label{fig:comparison_AB}
    \end{figure}

 We analyzed the light curves to check variability on short timescales, i.e., from hours to days. The light curves are shown in Figure \ref{fig:comparison_AB}, where the dashed lines represent 1$\sigma$ standard deviation. We calculated the $\chi^2$ and the degrees of freedom (d.o.f.) as a first approximation to test the variations. We considered the source to be variable if the count rate differed from the average above 3$\sigma$ (or 99.7$\%$ probability). To check the variability amplitude of the light curves, we calculated the normalized excess variance, $\sigma^2_{NXS}$. We followed prescriptions given by \cite{2003Vaughan} to estimate $\sigma^2_{NXS}$ and its error, err($\sigma^2_{NXS}$). We found that NGC\,3718 is not variable at $3\sigma$ confidence level. 
 The average count rates of the four segments and both detectors are listed in Table \ref{table:variability}. 
    
The \emph{XMM--Newton} observation was heavily affected by background flares, causing the loss of nearly half the exposure time. After removal of the flaring intervals, the 0.2--10.0 keV, background-subtracted light curve has an average count rate of 0.26 counts s$^{-1}$, and no significant variability over the length of the observation, 31 ks. The resulting excess variance is $\sigma^2_{NXS}=0.002\pm 0.003$, and therefore consistent with zero.

    \begin{table}
    \begin{center}
    
    \caption[lightcurve]
    {Statistics for the \emph{NuSTAR} light curves} 
        \tabcolsep=0.11cm
    \label{table:variability}
    \begin{tabular}{ccccccccc}
    \hline \hline
    Obs ID & FPM & $\sigma^2_{NXS}$ & err($\sigma^2_{NXS}$) & counts & error  \\
     &    &  & & (counts s$^{-1}$) & (counts s$^{-1}$) \\
    \hline 
    60301031002 & A &  	0.006 &  0.014 & 0.059 & 0.012  \\
                & B &	0.084 &  0.029	 & 0.061 & 0.013\\
    60301031004 & A &	0.023 &   0.012 & 0.047 & 0.011	 \\
                & B &       0.015 &   0.013 &0.046 & 0.012  \\
    60301031006 & A &		 0.041 &  0.015 & 0.056 & 0.012	 \\
                & B &      -0.0004 &  0.009 & 0.058 & 0.012  \\
    60301031008 & A &  	-0.014 &  0.014	 & 0.049 & 0.015 \\
                & B &      -0.014 &  0.033 & 0.035 & 0.015  \\
    \hline
    \end{tabular}
    \begin{minipage}{8cm}
    \textbf{Notes:} Obs ID (Col. 1), focal plane module (Col. 2), normalized excess variance with errors (Cols. 3 and 4), mean of counts and its errors (Cols. 5 and 6).
    \end{minipage}
    \label{tab:energetics}
    \end{center}
    \end{table}

    We also checked for spectral variability between the \nustar\ exposures by fitting each individual observation (for FPMA and FPMB) with an absorbed power-law with all parameters free. We found that NGC 3718 showed variations in the normalization smaller than 4\% and in $\Gamma$ smaller than 3\%, with values close to $\sim$1.94, on a timescale of ten days, all within the uncertainties in these parameters at $1\sigma$ level. The simultaneous fit therefore resulted in a non-variable spectrum on a 10-day timescale.

    Variations within the observed timescales were not detected between different spectra studied here. Therefore, the spectra of all the \nustar\ epochs were combined to increase the sensitivity, producing a single spectrum for each detector, A and B.  Since the  \emph{XMM--Newton} observation was carried out during one of the \nustar\ exposures,  the spectra from both observatories can be considered simultaneous.   We compared these contemporaneous \emph{XMM--Newton}/\nustar\ data sets in the same energy band (3--10 keV) to exclude calibration  differences between these instruments. We fitted the data with a power-law and fixed the slope, allowing variations in the normalization. We found values of $N_{XMM}=(5.1\pm0.2)\times 10^{-4}$ for \emph{XMM--Newton} and  $N_{Nustar}=(5.2\pm0.3)\times 10^{-4}$ for \nustar\ . Therefore, for this source, the instrument responses are consistent.

	\subsection{Spectral analysis}
    \label{sec:spectral_analysis}

    The spectral analysis of the \emph{NuSTAR}, \emph{XMM--Newton} and \emph{Swift}/BAT data was performed using {\sc XSpec} version 12.10.0 \citep{1996arnaud}. All the errors reported throughout the paper correspond to 90$\%$ confidence, unless otherwise noted. In this work we only used  \emph{XMM-Newton} observations from the EPIC-PN because of its higher throughput \citep{2001struder} and because inclusion of the EPIC-MOS spectra resulted in too much statistical weight to the low energy range data points compared to the \nustar\ and \emph{Swift}/BAT data. For all spectral fits, we included a multiplicative constant normalization between FPMA, FPMB, EPIC-PN and BAT to account for calibration uncertainties between the instruments and possible variations between the \emph{Swift}/BAT and \emph{Nustar} and \emph{XMM-Newton} exposures. We found that these calibration uncertainties are close to unity except in the case of \emph{Swift/}BAT data, where this constant is close to 4. This difference is shown in Table 3 and discussed in Section 4.1.
    
    NGC 3718 shows weak Fe K$\alpha$ emission. For this reason, we started our spectral analysis with a simple absorbed power-law with a high-energy cut-off (\texttt{phabs*zphabs*cutoffpl} in {\sc XSpec}. In this model the \texttt{phabs} component is associated with absorption from our Galaxy and fixed to 1.07$\times$10$^{\rm 20}$ cm$^{-2}$, obtained using the  $N_{H}$ tool within {\sc ftools} \citep{1990Dickey, 2005karberla}. The \texttt{zphabs} component is associated with absorption from the nuclear region. We note that the cut-off power-law model for the continuum emission is a phenomenological model that can represent Comptonized emission from a corona or ADAF or synchrotron emission from a jet, so the source of X-ray emission in the model can correspond to any of these structures. 

\begin{figure}
    \resizebox{\hsize}{!}{\includegraphics{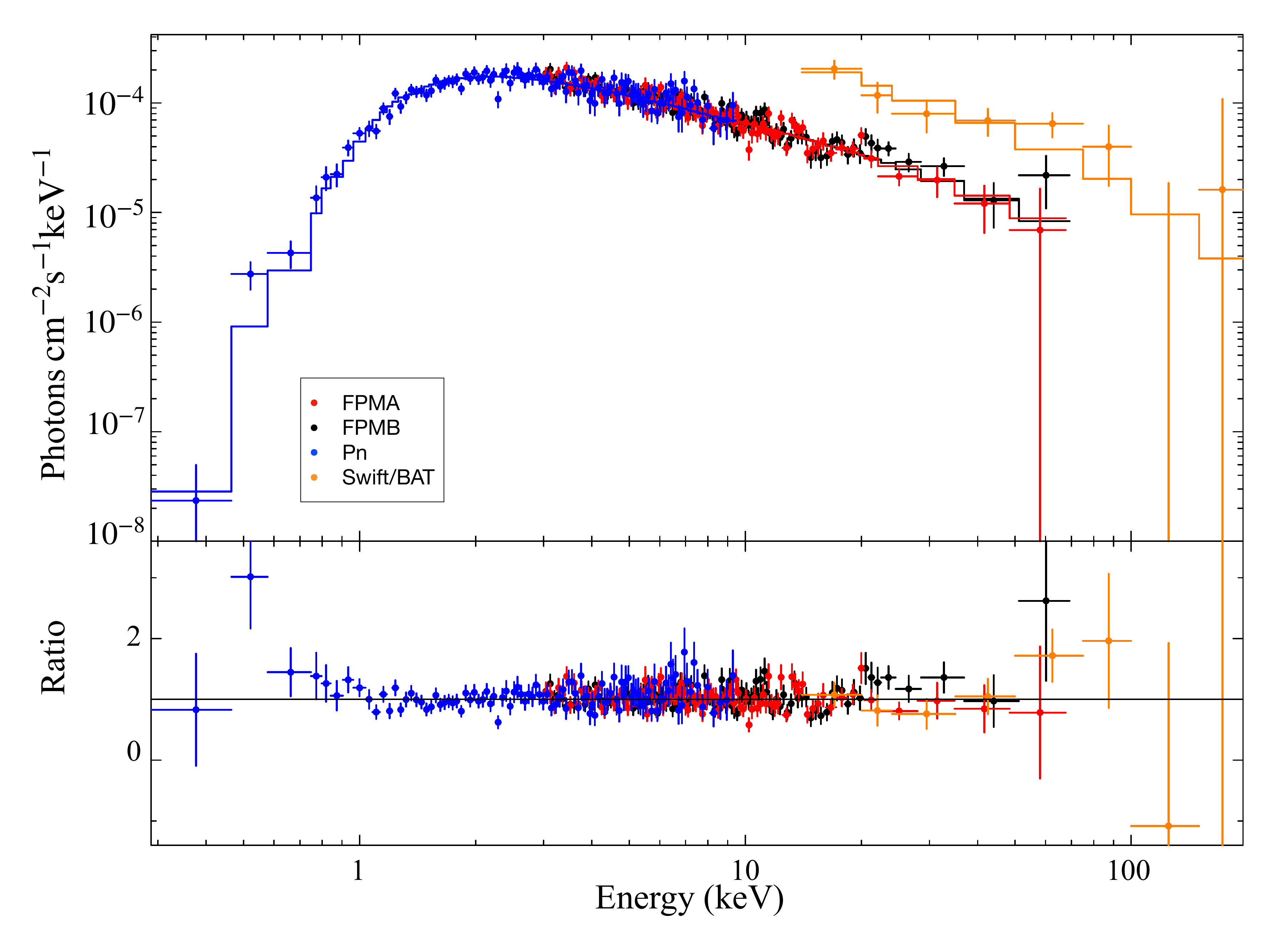}}
    \caption{Upper panel: Best-fit cut-off power-law model (solid line) to the \emph{NuSTAR} FPMA and FPMB, \emph{XMM-Newton} pn, and \emph{Swift}/BAT spectra of NGC\,3718 (filled circles). Lower panel: Residuals in terms of data-to-model ratio.
    }
     \label{fig:simple_model}
    \end{figure}

    This fit results in a $\chi^{2}$= 427.59 for 374 d.o.f. We find a significant intrinsic hydrogen column density of  $N_{H}$=(8.9$^{+0.7}_{-0.6}$)$\times$ 10$^{21}$ cm$^{-2}$, showing that the coronal emission is absorbed, $\Gamma$ = 1.78$\pm$0.08 and  $E_{\rm cut}$=73$^{+111}_{-30}$ keV. The data and best-fit cut-off power-law model are shown in Figure \ref{fig:simple_model}. This model fails to adequately fit the spectral continuum, leaving obvious structured residuals. 
    In order to improve the spectral fit, we studied the residuals in the soft (0.3--2.0 keV) and hard (2.0--110 keV) energy bands.

In order to interpret the parameters of the cut-off powerlaw in a physically meaningful way we replaced this component with the  comptonization model \texttt{CompTT}, which describes Comptonization of soft photons in a hot plasma \citep{titarchuk1994} to produce the primary continuum. We find the following constraints on the electron temperature (T$_e$) and optical depth of the corona ($\tau$): kT$_e=44^{+34}_{-24}$, $\tau=0.8^{+0.4}_{-0.8}$, where part of the uncertainty in these parameters is due to the degeneracy between them. In order to place tighter constraints on the roll-over energy, below we continue to use a phenomenological model for the continuum, i.e., a cut-off power-law. We note as well that these uncertainties do not incorporate the effect of a possible reflection component, which, as discussed below, can be important.

\subsection{Soft energy band}
\begin{figure}
    \resizebox{\hsize}{!}{\includegraphics{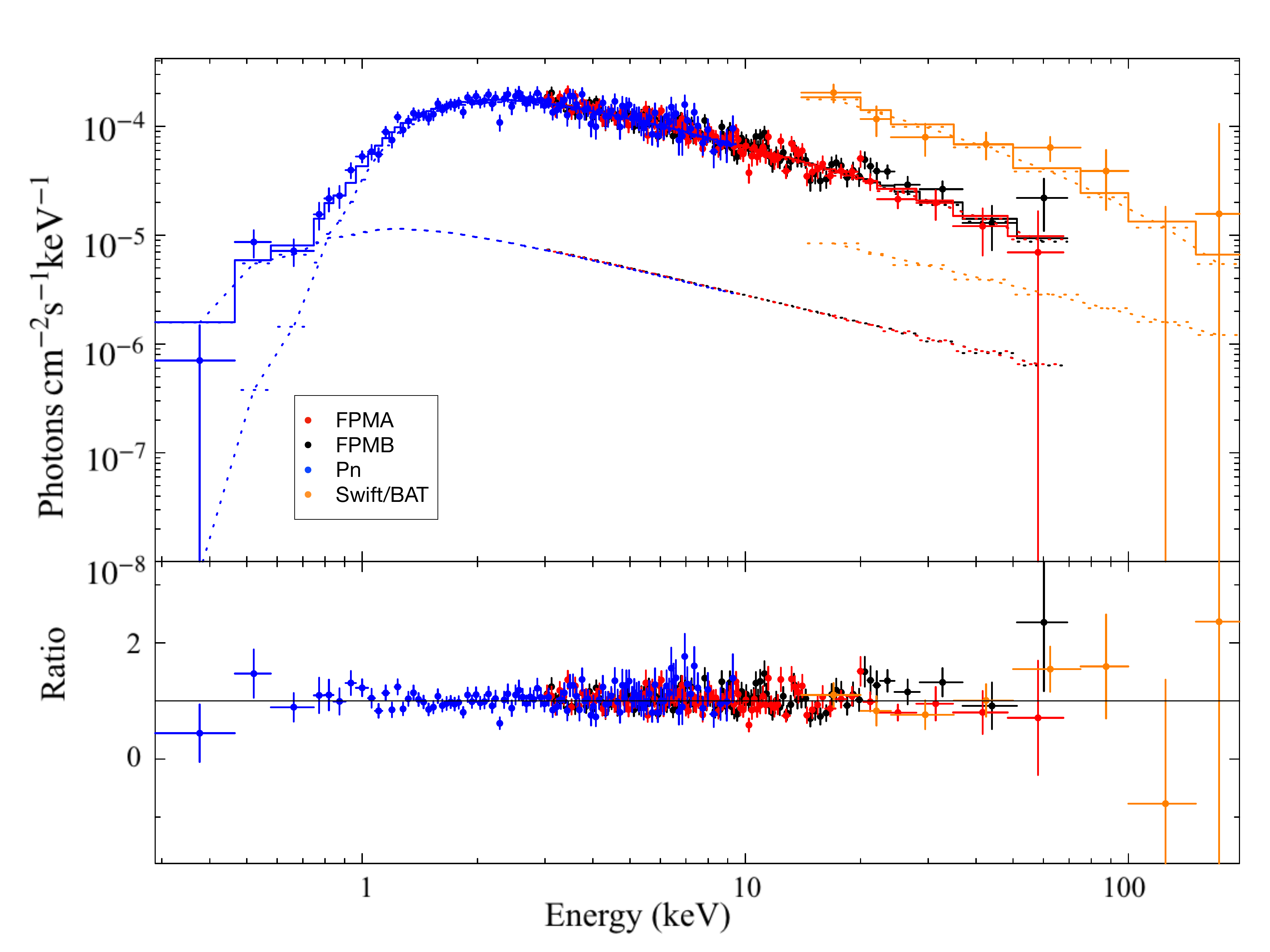}}
    \caption{Upper panel: Best-fit cut-off power-law + scattered model component (solid line) to the \emph{NuSTAR} FPMA and FPMB, \emph{XMM-Newton} pn, and \emph{Swift}/BAT spectra of NGC\,3718 (filled circles). Lower panel: Residuals in terms of data-to-model ratio.}
     \label{fig:simple_model_with_abs}
    \end{figure}

In the soft energy band, we added a thermal or scattered power-law component in order to improve the spectral fit. Adding a power-law component under a different host absorber, with a normalization of a few percent of the primary power-law with an identical slope, resulted in a good fit with $\chi^{2}$= 417.36 for 372 d.o.f., statistically better than the simpler model.  An alternative optically thin thermal component, modeled with {\sc APEC}, performed slightly worse. For this reason, we incorporate the scattered power-law component in the following models.  The improvement of the model including the scattered component is clearly seen in Figure \ref{fig:simple_model_with_abs}. Note that the intrinsic and scattered power-laws are very similar in shape, in the models they are mainly distinguished by the absorber that they have in front (the scattered component has a  lower column density than the intrinsic power-law emission). Leaving the relative normalization free sometimes results in an inversion of the components, which affects the correct identification of the absorbers. To make sure that the spectrum is not dominated by the scattered component at high energies, the relative normalization was restricted to a maximum of 4\% of the normalization of the nuclear power-law, where this particular value was obtained from a simple fit.

The column density of the absorber acting on the extended scattered power-law in this fit is $N_{{\rm H},S}=(2.1^{+1.0}_{-0.1})\times 10^{21}$ cm$^{-2}$, and the absorber acting on the nuclear power-law is $N_{{\rm H},H}=(1.0 \pm 0.1)\times 10^{22}$ cm$^{-2}$, $\Gamma$ = 1.84$^{+0.08}_{-0.06}$ and E$_{\rm cut}=99^{+153}_{-50}$ keV. A similar model was shown by \cite{2009Gonzalez} to be a good representation of the X-ray spectrum of 82 LINERS in the 0.2--10 keV band with \emph{XMM-Newton} and \emph{Chandra} data.  

\subsection{Hard energy band}

\begin{figure}
    \resizebox{\hsize}{!}{\includegraphics{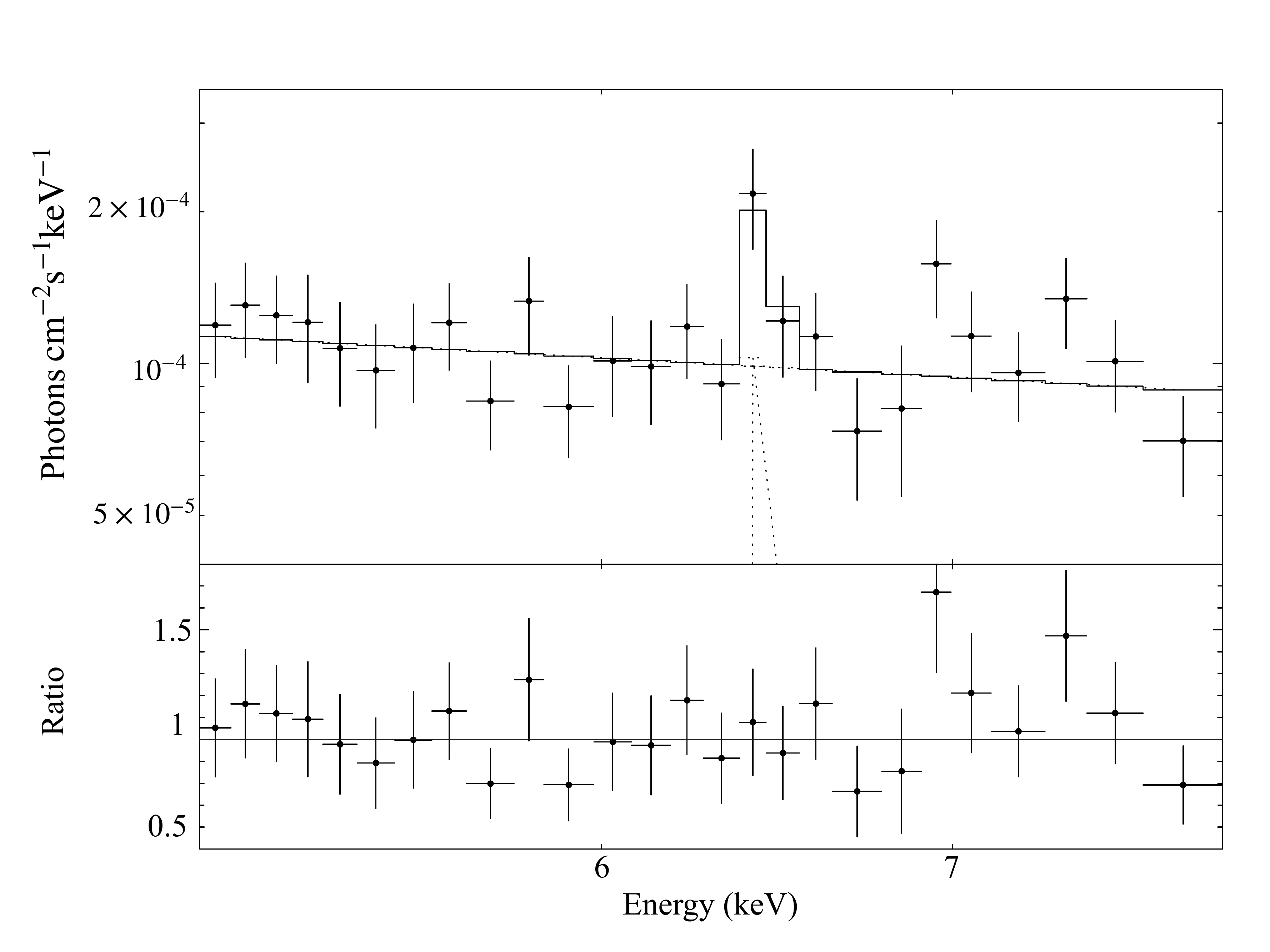}}
    \caption{Upper panel: Count spectrum of the \emph{XMM-Newton} observation zoomed-in at the Fe complex. Lower panel: Residuals in terms of data-to-model ratio.}
     \label{fig:iron_emission}
    \end{figure}

The small residuals observed in the hard energy band could be an effect of reflection from an accretion disc or a torus. The effect of this component is thought to affect the curvature of the hard X-ray emission and is responsible for the creation of the iron emission line at 6.4 keV \citep{1990Pounds, 1994Nandra}.
As a first approach to quantify possible reflection features, we investigated the presence of iron emission in the spectrum of NGC\,3718. We analyzed data in the energy range 5.0--8.0 keV to constrain the need of a narrow Gaussian component. First, we fit a power-law model in this range and obtain $\chi^{2}$/d.o.f=20.27/24=0.84. Then we add a Gaussian component to study the improvement of the fit, and found $\chi^{2}$/d.o.f=16.66/22=0.76. This component has a line centered at 6.5$^{+0.1}_{-0.2}$ keV, consistent with FeK$\alpha$ (6.4 keV) emission and equivalent width EW = 0.11$^{+0.04}_{-0.05}$ keV.  Note that we fixed the width of the emission line to 0.01 keV, below the instrumental resolution. Figure \ref{fig:iron_emission} shows the best-fit model in a zoom-in at 5.0-8.0 keV with a Gaussian fit and its residuals. This figure shows that, although small, an emission line at 6.4 keV is consistent with the spectrum of NGC\,3718.

\begin{figure}
    \resizebox{\hsize}{!}{\includegraphics{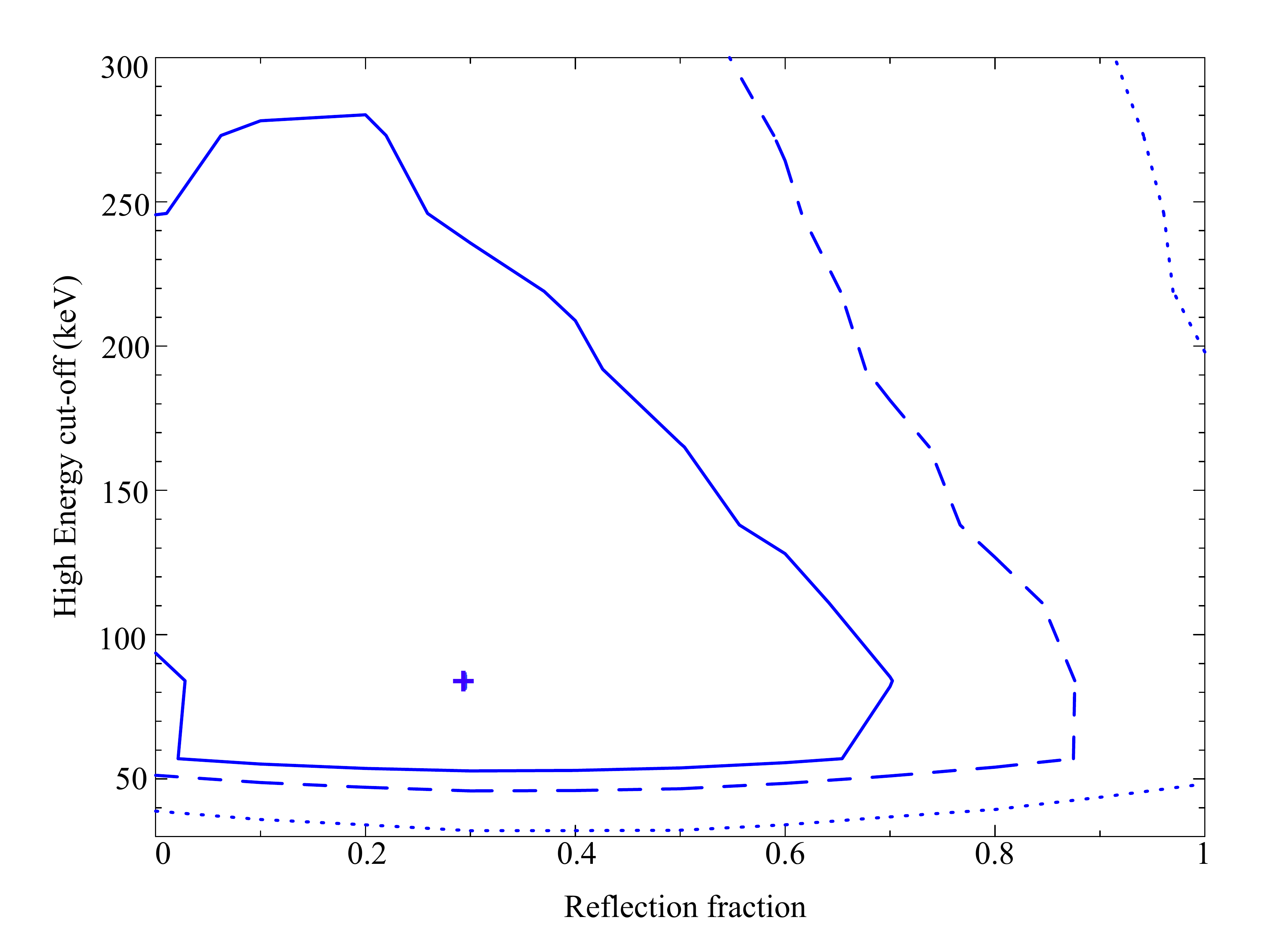}}
    \caption{Two-dimensional $\Delta \chi^{2}$ contours for reflection fraction and cut-off energies for NGC\,3718 with \texttt{pexmon} reflection model. The solid blue line represents the 1$\sigma$, the dashed line 2 $\sigma$ level and the pointed line 3$\sigma$ contour level. The best-fitting values are marked by a + sign. }
     \label{fig:conto_pexmon}
    \end{figure}

In order to assess the reflection fraction limit that we can derive from our data, we firstly used a simple reflection model. We used the \texttt{pexmon} \citep{2007nandra} model implemented in {\sc XSpec} which uses as continuum an exponentially cut-off power-law. Since the pexmon model represents both the reflected and intrinsic emission, the model employed was pexmon + scattered component, removing the coronal cut-off power-law. 
In this way, the model parameter  $R_{f}$ corresponds to the reflection fraction and is a free parameter. We found a good fit to the data with this model with $\chi^{2}$=416.43 for 371 d.o.f, with a best-fitting $\Gamma$=1.85$\pm$ 0.08,  $E_{\rm cut}$=84$^{+93}_{-39}$ keV and reflection fraction  $R_{f}$<0.67 (with the best-fitting value of 0.30). The reflection fraction is partially anti-correlated with the cut-off energy as can be seen in the contour plot in Figure \ref{fig:conto_pexmon}; lower reflection fractions allow higher cut-off energies, although this parameter is still constrained at the $1 \sigma$ level. 
 
The improvement on the fit including \texttt{Pexmon} is   marginal and an F-test shows that it is not statistically significant.  The accreting black hole, however, cannot be completely isolated, there must be material around it, such as the accretion flow itself and a small BLR \citep{2018cazzoli}. In agreement with this assumption, our previous analysis shows that the spectrum is well described when a reflection component is included, although it is formally also consistent with no reflection, perhaps due to limited signal to noise ratio. Despite the weakness of the possible reflection, it should not be ignored, because if present it will produce curvature in the hard X-ray spectrum that would otherwise be misinterpreted as an intrinsic rollover of the primary continuum.  Furthermore, our data (among the deepest available hard X-ray spectrum for a LLAGN) allow us to put physically meaningful limits on the amount and distribution of the material surrounding the AGN. Therefore, in the following discussion we model a reflection component, without a priori restrictions on its strength. 

Aiming at fitting the data with the most representative physical model, we studied reflection models which might come from a neutral reflector as modeled by \texttt{Borus02}  \citep{2018balokovic} or \texttt{MYTorus} \citep{2009Murphy}, as well as from an ionized accretion disc (\texttt{Relxill}, \citealt{2013garcia}). The final model employed in the analysis is defined as:
\vskip 0.3cm

\boxed{ C \times  N_{\rm Gal} (N_{{\rm H},S} \times  $ PL $ +  N_{{\rm H},H}\times  $ cPL $ +  N_{{\rm H},H} \times  {table})} \vskip 0.5cm

\noindent where $C$ represents the cross-calibration constant between different instruments and  $N_{\rm Gal}$ is the Galactic absorption (\texttt{phabs} in {\sc XSpec}); $N_{{\rm H},S}$ is the column density of absorbing material acting on the scattered power-law, PL is the power-law of the scattered component;   $N_{{\rm H},H}$ is the absorbing material that acts on the nuclear components (power-law and torus or disc reflection)\footnote{We made a test with an absorber acting or not in the torus-like reflector and we found that it is indistinguishable. The same result was found for the absorber on the scattered component.}; cPL is a cut-off power-law (\texttt{cutoffpl} in {\sc XSpec}) representing the primary X-ray emission and ``table'' represents the  different reflection models that will be used.  To be consistent, we ignore data below 0.7 keV in the following analysis because the \texttt{MYTorus} model only works above that energy. We therefore fixed the parameters of the scattered power-law component to the values that were previously obtained.

In the following, we compare the effect of the different reflectors on the resulting primary continuum parameters (i.e., $\Gamma$ and  $E_{\rm cut}$) and study the properties of the reflector itself. Elemental abundances are assumed to be solar for all models. 

   \subsubsection{Torus reflection model: Borus}

\citealt{2018balokovic} developed a radiative transfer code that calculates the reprocessed continuum of photons that are propagated through a cold, neutral and static medium. In this work, we used the geometry that corresponds to a smooth spherical distribution of neutral gas, with conical cavities along the polar directions (\texttt{Borus02}). The opening angle of the cavities, as well as the column density and the inclination of the torus, are free parameters. The reflected spectrum of this torus is calculated for a cut-off power-law illuminating continuum, where  $E_{\rm cut}$, $\Gamma$ and normalization are free parameters. Therefore, combining \texttt{Borus02} with a cut-off power-law with parameters tied to those of the \texttt{Borus02} illuminating source, a consistent model can be obtained.  We tied the opening angle, $\theta_{tor}$, to the inclination angle, $\theta_{incl}$, to ensure a direct view of the central engine, and modeled the direct coronal emission separately with a cut-off power-law under a neutral absorber with \texttt{zphabs}. We recall that $\theta_{tor}$=0 corresponds to a pole-on view. The free parameters in this model are the column densities along the line-of-sight, the covering factor and column density of the reflector, $\Gamma$ of the primary emission, its  $E_{\rm cut}$ and normalization, which in turn is tied to the normalization of the reflector. 

\begin{figure}
    \resizebox{\hsize}{!}{\includegraphics{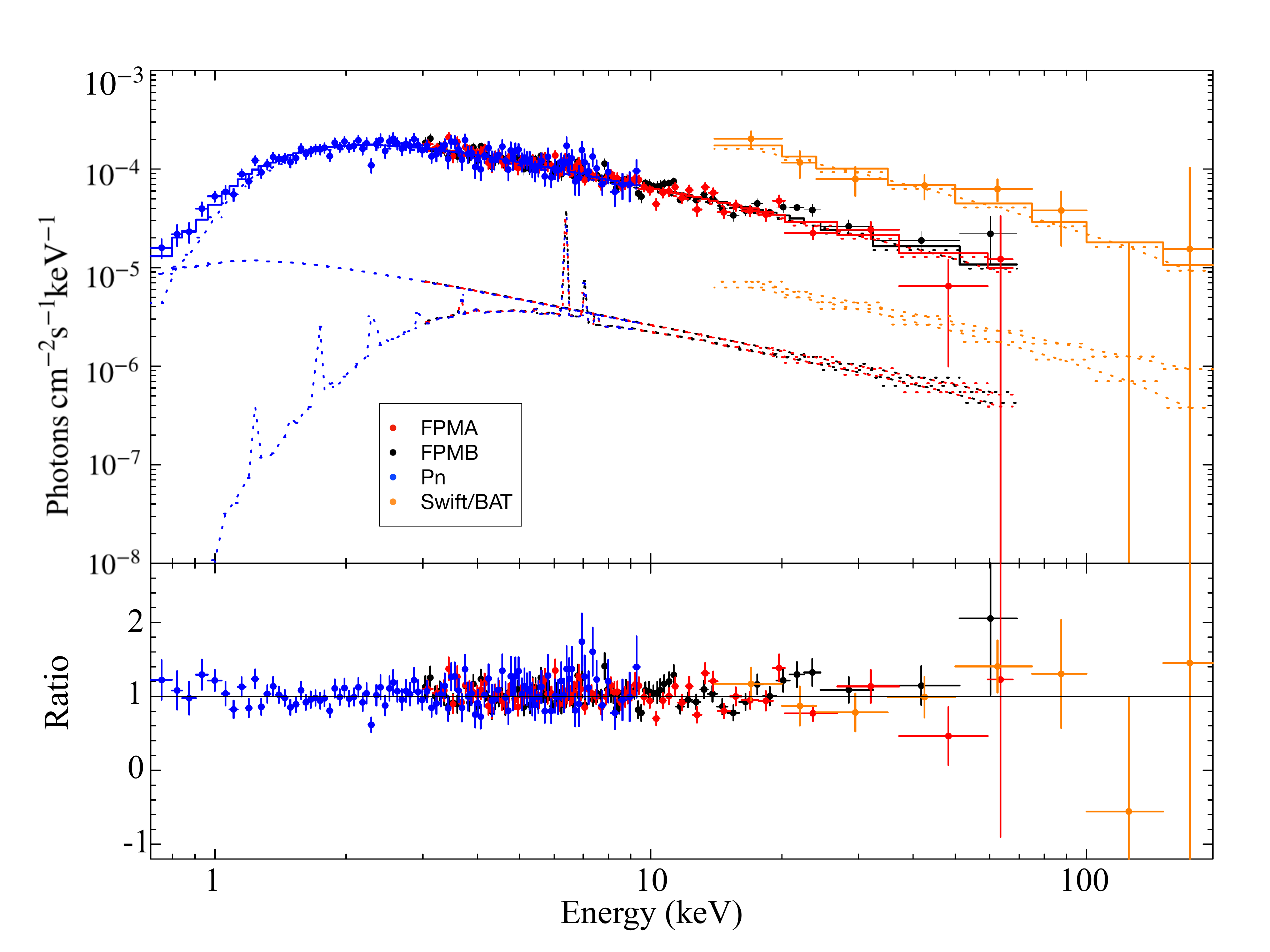}}
    \caption{Upper panel: Best-fit \texttt{Borus+cut-off PL} model (solid line) to the \emph{NuSTAR} FPMA and FPMB, \emph{XMM-Newton} pn, and \emph{Swift}/BAT spectra of NGC\,3718 (filled circles). Lower panel: Residuals in terms of data-to-model ratio.}
     \label{fig:model_torus_bat}
    \end{figure}

The best-fit model is shown in Figure \ref{fig:model_torus_bat}. This fit is statistically acceptable with $\chi^{2}$= 405.56 for 364 d.o.f. and no clear structure in the residuals.  The best-fitting values for the $\Gamma$,  $E_{\rm cut}$, and absorption can be found in Table \ref{tab:resumen}.   

\begin{figure}
    \resizebox{\hsize}{!}{\includegraphics{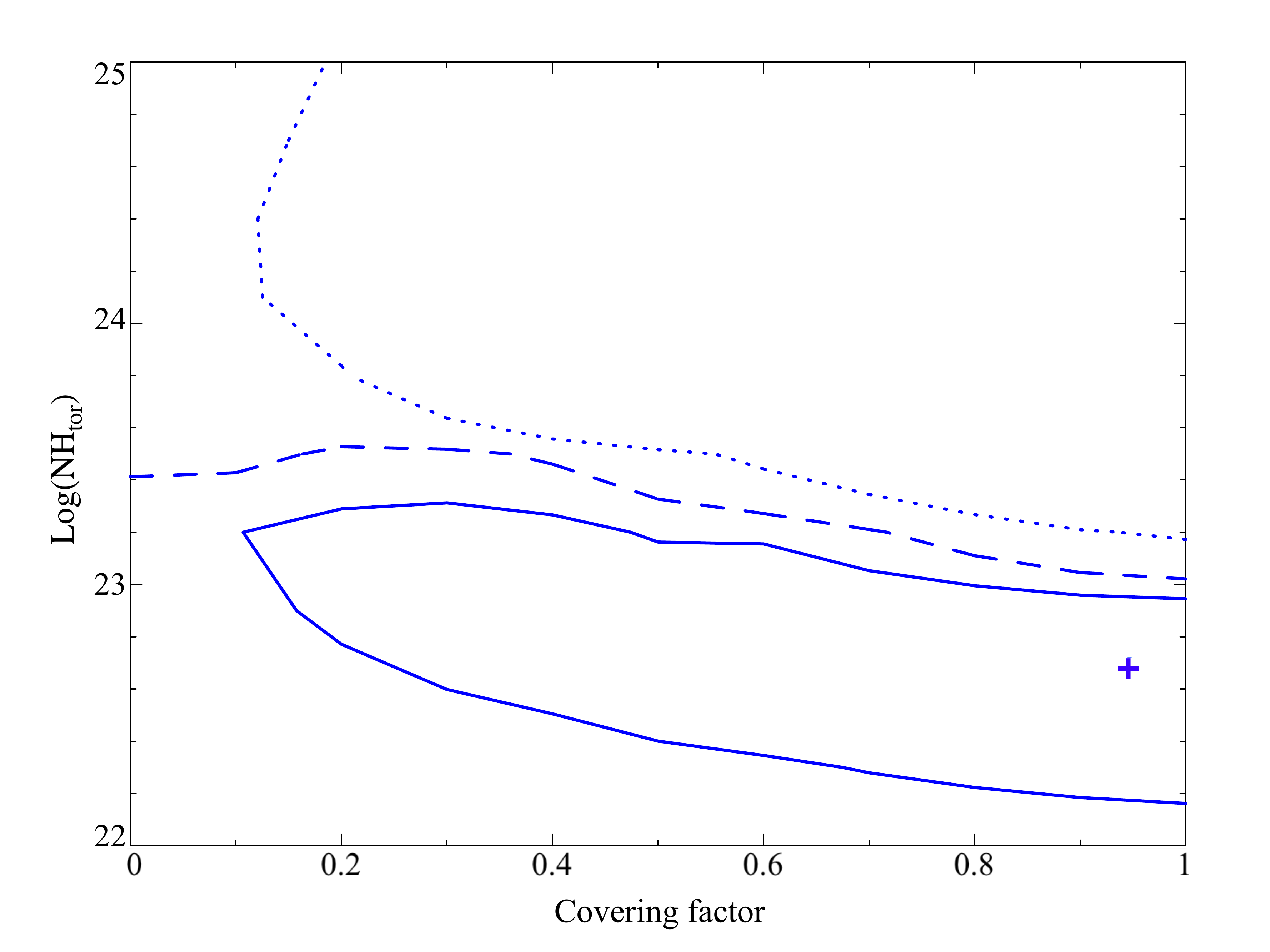}}
    \caption{Two-dimensional $\Delta \chi^{2}$ contours for torus column density and covering factor for NGC\,3718, modelled with \texttt{Borus}. The solid, dashed and dotted contours denote 1$\sigma$,2$\sigma$ and 3$\sigma$ respectively. The "+" symbol represents the best fit value. The reflector is constrained to have low equatorial column density  $N_{H}$ $<10^{23.0}$ cm$^{-2}$ with a large best-fitting covering fraction of 0.95 of the sky, although smaller covering fractions down to 0.1 are permitted at the 1$\sigma$ level.}
    \label{fig:BORUS_cont1}
    \end{figure}

We also put constraints on the column density of the reflector and its covering fraction.  As can be seen in the contour plot in Figure \ref{fig:BORUS_cont1}, the parameter space allowed by the data is broad. The reflector is only constrained to have a relatively low equatorial column density $N_{H} < 10^{23.2}$ cm$^{\rm -2}$ for any covering fraction, at the 1$\sigma$ level, and $N_{\rm H} < 10^{23.5}$ cm$^{\rm -2}$ at the 2$\sigma$ level. The existing reflection features only require a contribution from the torus in this model at the 1$\sigma$ level, with a column density $N_{\rm H} > 1.6\times10^{22}$ cm$^{\rm -2}$.

We note that this low column density and high covering fraction solution is consistent with the absorption seen along the line of sight to the corona ($N_{{\rm H},H}\sim 10^{22} \rm {cm}^{-2}$, see Figure \ref{tab:resumen}), so a model where a Compton thin torus covers a high fraction of the sky 
is consistent with the data both in terms of reflection and absorption.

 \subsubsection{Torus reflection model: MYtorus}

The \texttt{MYtorus} \citep{2009Murphy} model proposes a toroidal geometry where the covering fraction is fixed to 0.5, although different values can be mimicked by varying the normalization of the torus relative to the direct coronal emission, as we will do here. The equatorial column density and inclination angle are free parameters. In this model we fix: the metallicity to solar by tying the normalization of the scattered and fluorescent FeK$\alpha$ line components, and the foreground Galactic absorbing column density. We include a cut-off power-law under a neutral absorber as before, to model the direct coronal emission. The model is the same that was used with \texttt{Borus}, but replacing the reflector with the \texttt{MYTorus} tables. The free parameters in this model are the absorbing column densities along the line of sight, the column density and normalization of the scattered component, the normalization of the power-law, its  $E_{\rm cut}$ and its $\Gamma$. 
 
   The high energy cut-off is not a parameter for the illuminating source in this model, but a few tables exist for different input termination energies. We repeated the spectral fits with the available tables, calculated for illuminating continua with termination energies of 100, 160, 200, 300 and 500 keV to search for a solution where the cut-off energy of the power-law is below the termination energy of the reflection, as otherwise the model would not be consistent. 
   
\begin{figure}
    \resizebox{\hsize}{!}{\includegraphics{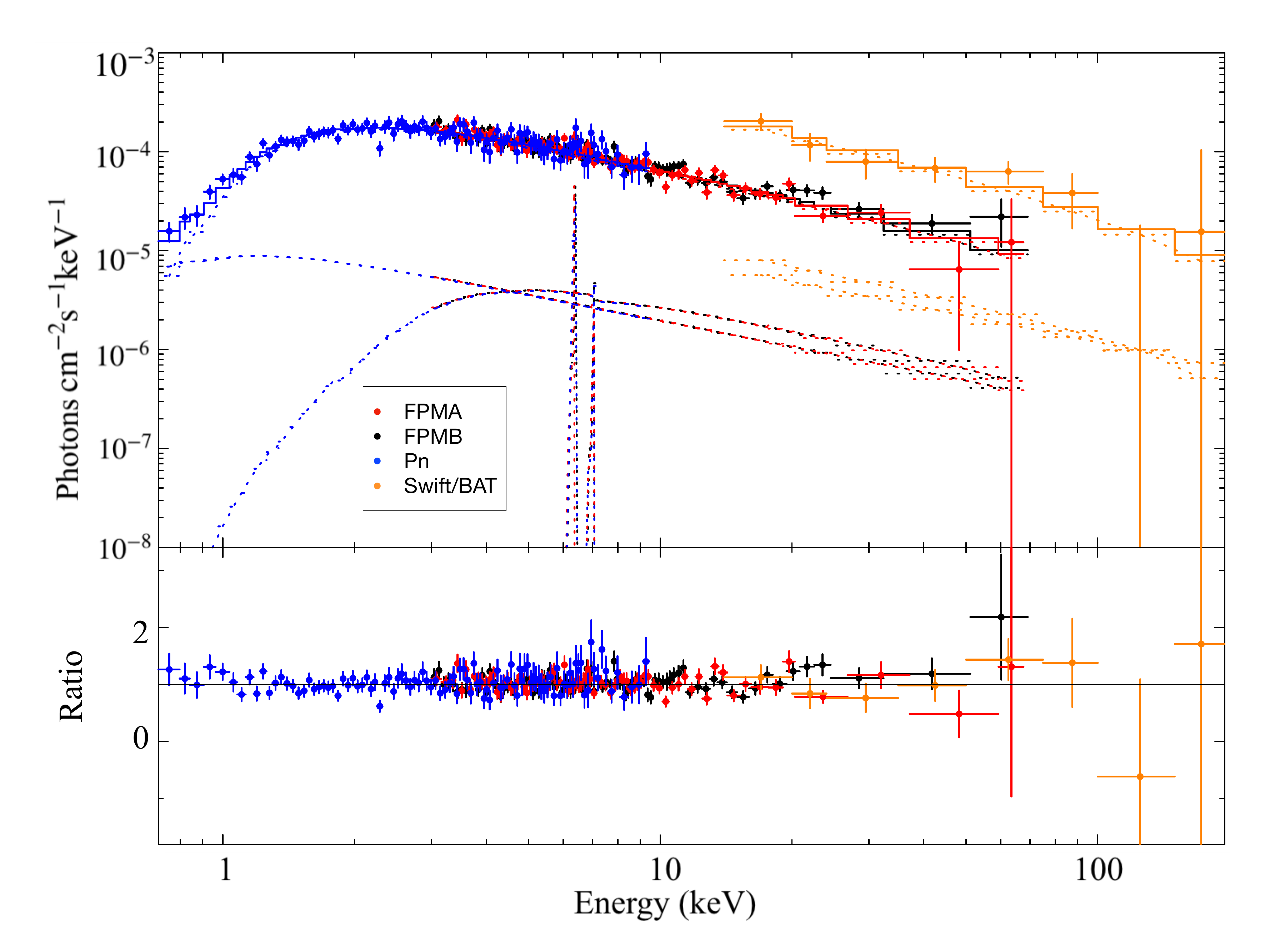}}
    \caption{Upper panel: Best-fit \texttt{MYTorus+cut-off PL} model (solid line) to the \emph{NuSTAR} FPMA and FPMB, \emph{XMM-Newton} pn, and \emph{Swift}/BAT spectra of NGC\,3718 (filled circles). Lower panel: Residuals in terms of data-to-model ratio. }
     \label{fig:model_mytorus}
    \end{figure}

   In this work we will use the reflection model with termination energy at 300 keV since this is the lowest value of  $E_{\rm cut}$ that satisfies the previous condition. The best-fitting model is plotted in Figure \ref{fig:model_mytorus}. This fit is statistically acceptable with $\chi^{2}$=406.58 for 364 d.o.f. and has an equatorial column density of the reflector of 7.2$^{+6.4}_{-5.9} \times$ 10$^{22}$ cm$^{\rm -2}$. The absorbing column densities can be found in Table \ref{tab:resumen}.

 The covering fraction can be studied by comparing the normalization of the scattered component with the normalization of the power-law. The column density of the reflector and covering fraction follow the same  trend observed with \texttt{Borus02}, i.e., a low column density reflector (N$_{H}\leq$3$\times$10$^{22}$ cm$^{-2}$) with any covering fraction at 1$\sigma$ level, although a high covering fraction of the sky is preferred. 

\begin{figure}
    \resizebox{\hsize}{!}{\includegraphics{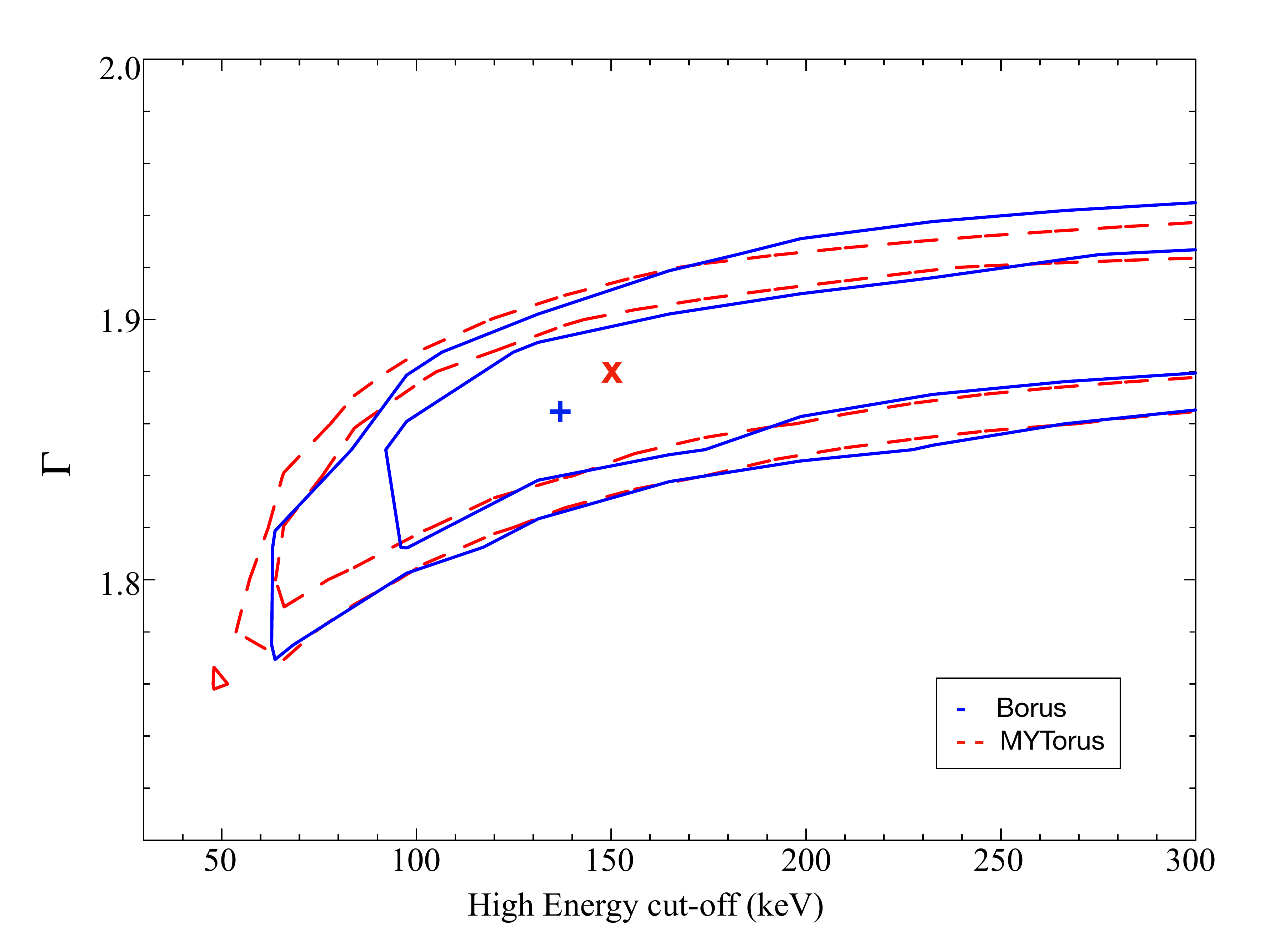}}
    \caption{Two-dimensional $\Delta \chi^{2}$ contours for $\Gamma$ and cut-off energies for NGC\,3718 for a neutral material as a reflector. Red dashed lines show the 1$\sigma$ and 2$\sigma$ levels for \texttt{MYTorus} and the "x" symbol represents the best fit values. Blue solid lines the 1$\sigma$ and 2$\sigma$ levels for \texttt{Borus02} and the "+" symbol represents the best fit values. Both models are in agreement. \texttt{Borus02} model starts in $\sim$100 keV at 1$\sigma$ level and \texttt{MYTorus} covers from $\sim$80 keV.} 
     \label{fig:cont_mytorus}
    \end{figure}

The contour plots for the $\Gamma$ and  $E_{\rm cut}$ for the \texttt{Borus02} and \texttt{MYTorus} reflection models are shown in Figure \ref{fig:cont_mytorus}. Both models are highly consistent, showing a range of allowed values of $\Gamma$ (i.e., 1.80-1.90). The main difference between them is the broader range in  $E_{\rm cut}$ obtained with 
\texttt{MYTorus}, reaching much lower values at $1\sigma$, but with similar $2\sigma$ level contours.  

 \subsubsection{Disk reflection models}
 
  \texttt{MYTorus} and \texttt{Borus} both model a neutral reflector with toroidal geometry. Another alternative is that reflection arises from an accretion disc. Thus we now  put constraints on the properties of a disc-like reflector and estimate the parameters of the primary emission.   

We explore ionized accretion disc reflectors using \texttt{Relxill} reflection models \citep{2013garcia}. This model calculates the reflected spectrum from the surface of an X-ray illuminated accretion disc by solving the equations of radiative transfer, energy balance, and ionization equilibrium in a Compton-thick and plane-parallel medium. \texttt{Relxill} is composed of many models; in this work we consider the case where the coronal spectrum 
is either a power-law with an exponential cut-off (\texttt{Xillver}) or a thermalized Compton spectrum (\texttt{XillverCp}). In the former case, the spectrum is described by the photon index ($\Gamma$) and the high energy cut-off ( $E_{\rm cut}$). In the latter case, the spectrum is described by $\Gamma$ and the electron temperature of the corona ($kT_{e}$). The other common parameter is the ionization, described by the ionization parameter ($\xi$), defined as the incident flux (F) divided by the density of the disc (n): $\xi=4\rm \pi$F/n [erg cm s$^{-1}$].  A low value implies that the disc is neutral. For increasing ionization parameter, the number and strength of the emission lines observed in the spectra generally decreases, leading to a fully ionized disc, which acts almost as a mirror and therefore the spectrum exhibits no line features (see \citealt{2013garcia}, for a more detailed description). In this model, this parameter is described by $\log (\xi)$ ranging from 0 for a neutral disc to 4.7 for a heavily ionized disc. 

Other important parameters are the iron abundance A$_{Fe}$ relative to the solar value (assumed to be solar in this work\footnote{We varied these parameters and found minimal effects on our results. }), redshift, and inclination. This model contains both the direct spectrum of the corona and the reflection spectrum.  
 We removed the coronal power-law from the model setup and replaced it, together with the reflected component, with either \texttt{Xillver} or \texttt{XillverCp} with a positive reflection fraction. In this way, the model parameter  $R_{f}$ corresponds to the relative fraction of coronal photons hitting the disc to those escaping to infinity. Our model in these cases is defined as:\vskip 0.5cm
 
 \boxed{C \times  N_{\rm Gal} (N_{{\rm H},S} \times $ PL $ +  N_{{\rm H},H} \times$table$)} \vskip 0.5cm

\noindent where ``table'' represents the accretion disc reflection model. We made a test using the \texttt{Xillver} model with a neutral accretion disc ($\log \xi=0$), leaving the reflection fraction as a free parameter, in order to study consistency with the results previously found with \texttt{pexmon}. We found that our result is compatible with \texttt{pexmon}, observing that the best fitting values of  $E_{\rm cut}$ and $\Gamma$ are almost identical. 

\begin{figure}
    \resizebox{\hsize}{!}{\includegraphics{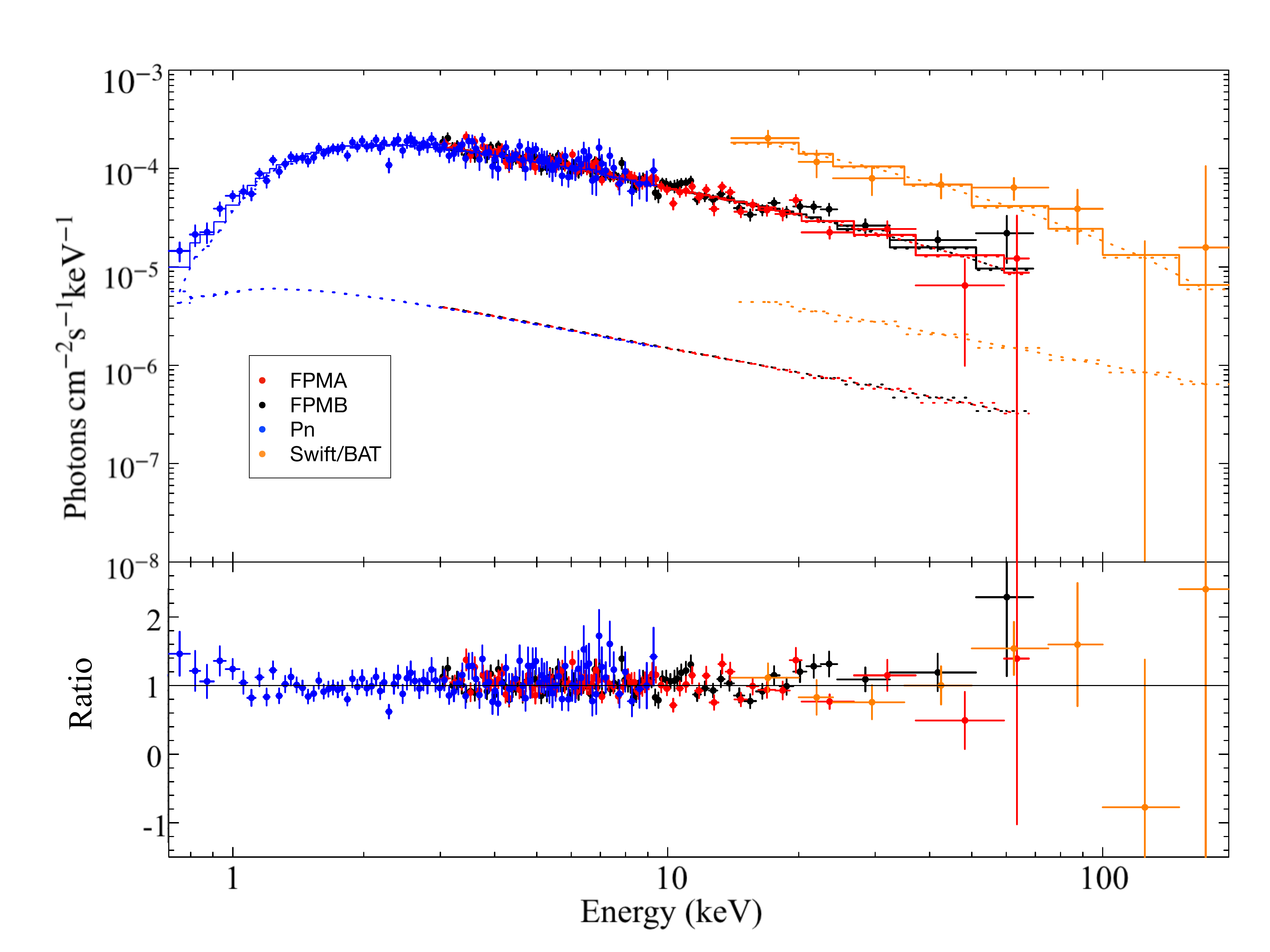}}
    \caption{Upper panel: Best-fit \texttt{Xillver} model (solid line) to the \emph{NuSTAR} FPMA and FPMB, \emph{XMM-Newton} pn, and \emph{Swift}/BAT spectra of NGC\,3718 (filled circles). Lower panel: Residuals in terms of data-to-model ratio.}
     \label{fig:xilver_model_spectrum}
    \end{figure}

\begin{figure}
    \resizebox{\hsize}{!}{\includegraphics{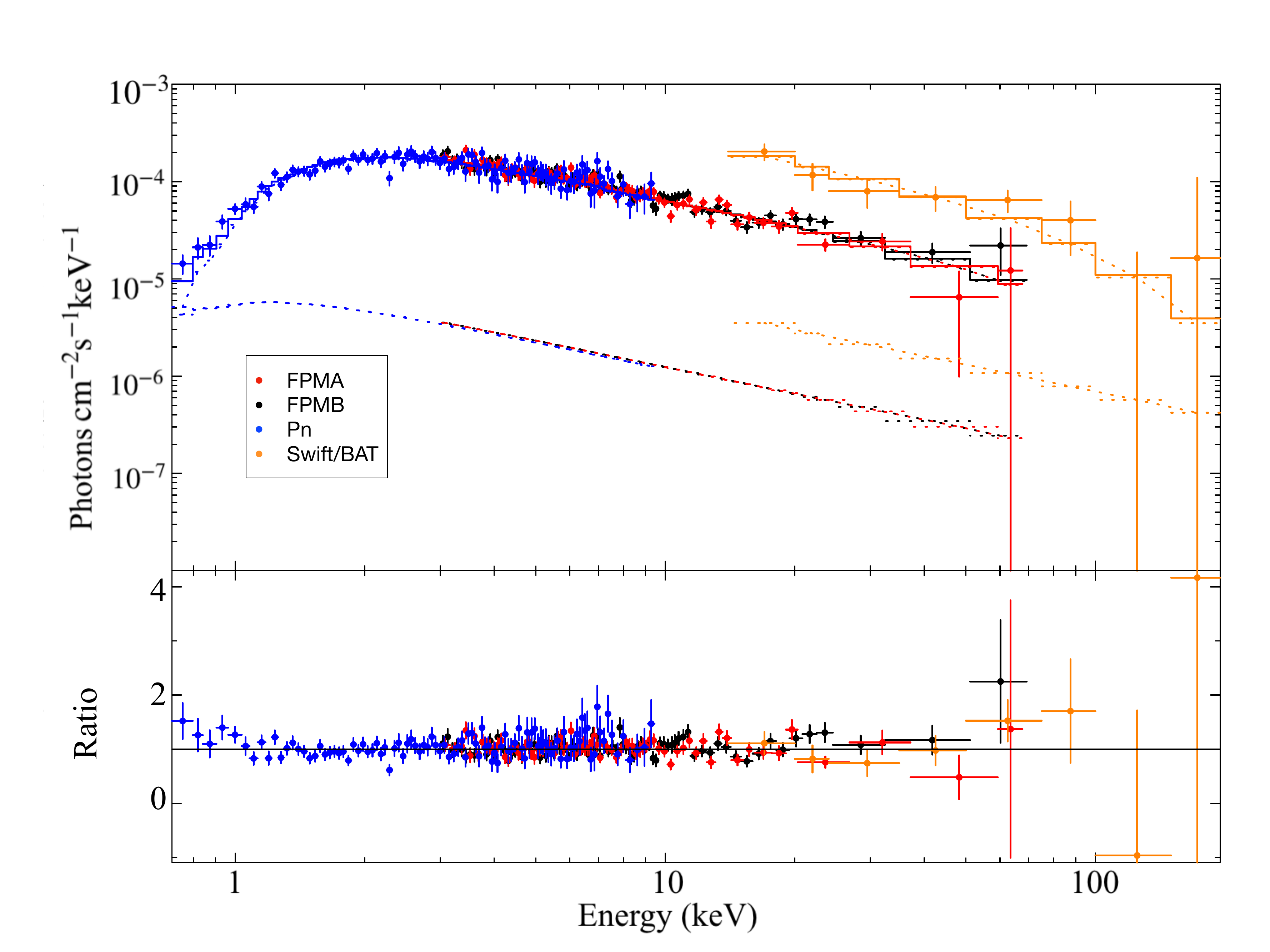}}
    \caption{Upper panel: Best-fit \texttt{XillverCp} model (solid line) to the \emph{NuSTAR} FPMA and FPMB, \emph{XMM-Newton pn, and \emph{Swift}/BAT} spectra of NGC\,3718 (filled circles). Lower panel: Residuals in terms of data-to-model ratio.}
     \label{fig:xilvercp_model}
    \end{figure}

Allowing the \texttt{Xillver} ionization parameter to vary, we find a good fit to the data, with $\chi^{2}=$ 409.59  for 363 d.o.f. The flat residuals suggest that all features in the data are fitted by this solution.  Replacing \texttt{Xillver} with \texttt{XillverCp} results in $\chi^{2}$= 407.21 for 363 d.o.f. and similar residuals. The best value for $\Gamma$,  $E_{\rm cut}$ and  $N_{{\rm H},H}$ of both models are presented in Table 3. The best-fitting \texttt{Xillver} model is shown in Figure \ref{fig:xilver_model_spectrum} and \texttt{XillverCp} in Fig \ref{fig:xilvercp_model}.

\begin{figure}
    \resizebox{\hsize}{!}{\includegraphics{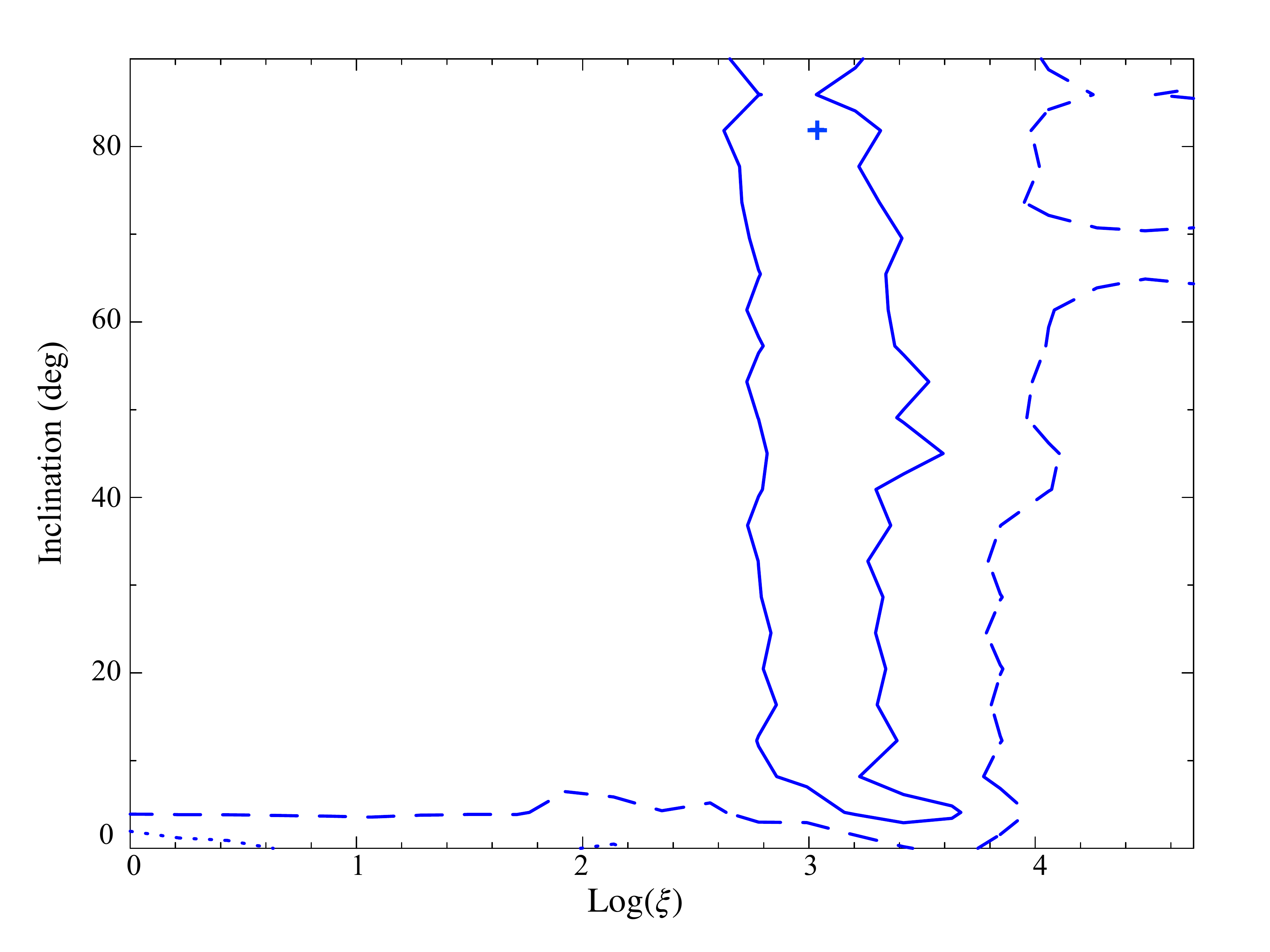}}
    \caption{Two-dimensional $\Delta \chi^{2}$ contours for ionization of the disc and inclination of the disc for NGC\,3718 for fits with \texttt{Xillver}. The blue solid, dashed and dotted contours represent the 1$\sigma$, 2$\sigma$ and 3$\sigma$ levels, respectively. The best value of $\log\xi$ is restricted to values between 2.8 and 3.4, while the inclination is unconstrained preferring high inclination ($\sim$80 degrees. }

     \label{fig:xilver_cont2}
    \end{figure}

Figure \ref{fig:xilver_cont2} shows contour plots of the disc inclination relative to the line of sight and the ionization degree of the disc for \texttt{Xillver}. We found that the inclination  is unconstrained for all the model, with the best value reaching $\sim$80 degrees. The ionization of the disc (log$\xi$) is well restricted to values between 2.8 and 3.5, with the best value $\log (\xi)\sim 3.1$. Replacing \texttt{Xillver} with \texttt{XillverCp}, the model produces the same result for these parameters.

    \begin{figure}
    \resizebox{\hsize}{!}{\includegraphics{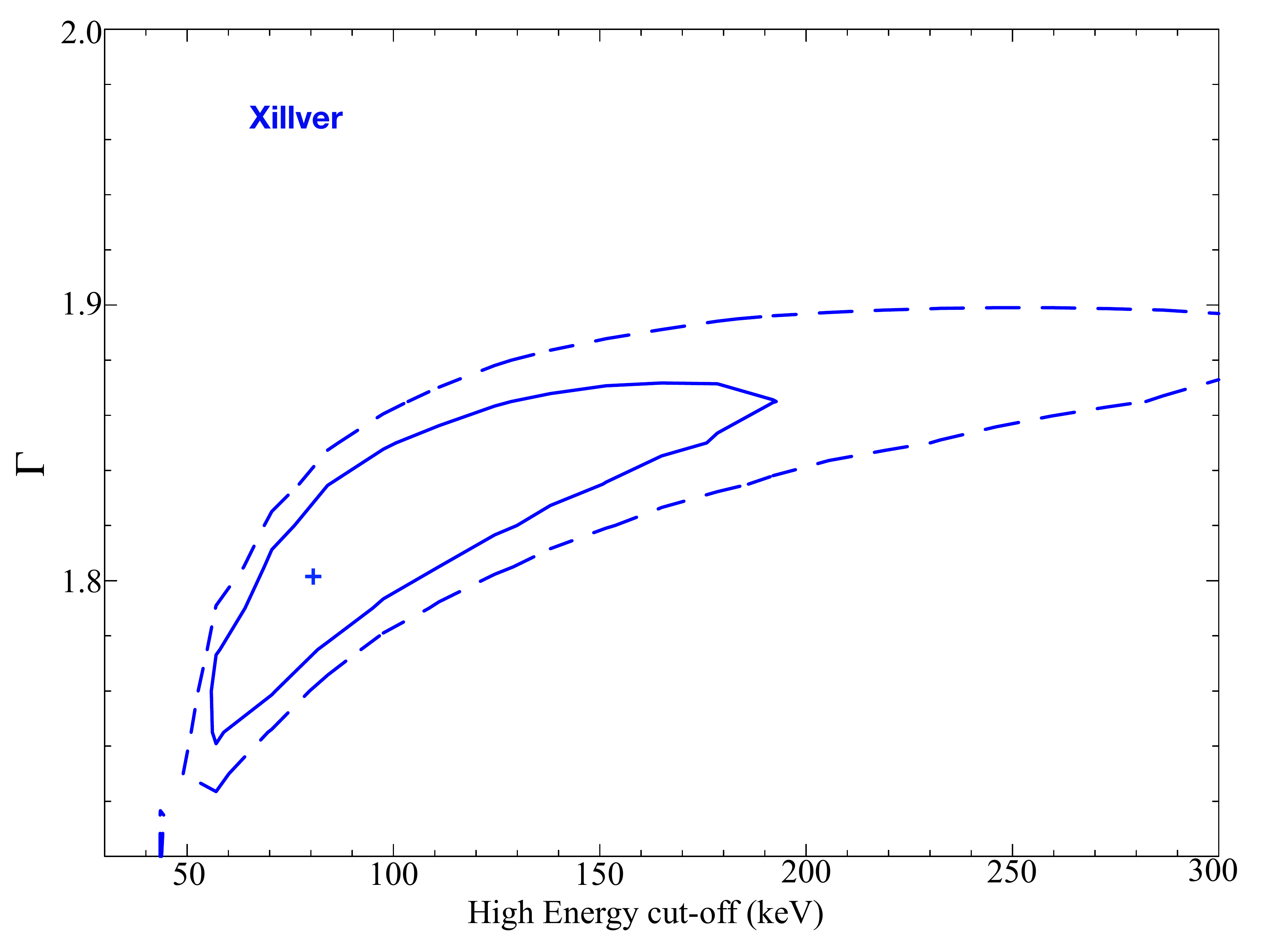}}
    \resizebox{\hsize}{!}{\includegraphics{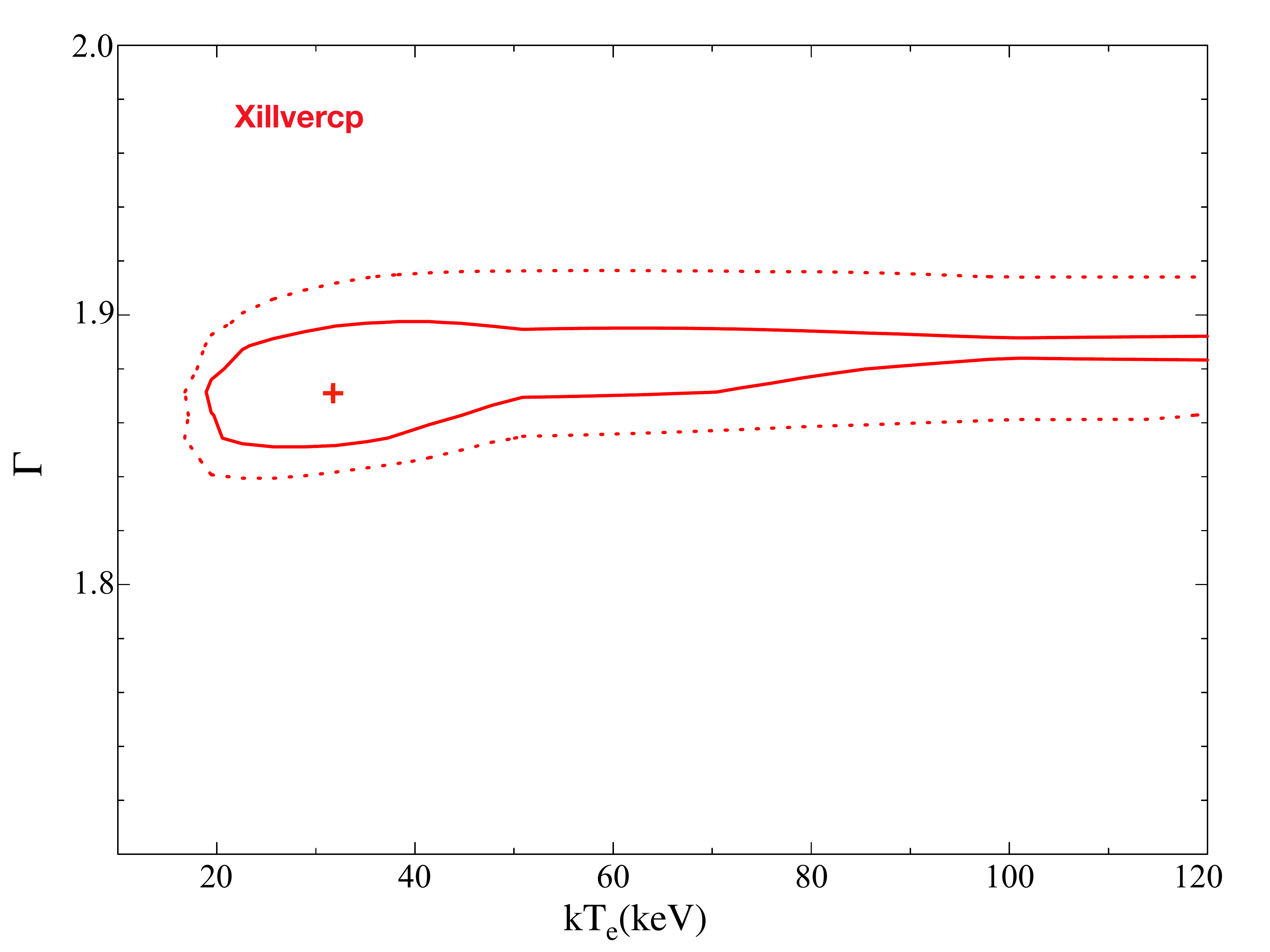}}
    \caption{Two-dimensional $\Delta \chi^{2}$ contours for $\Gamma$ and cut-off energy for NGC\,3718. The solid contour is the 1$\sigma$ level and the dashed contour the 2$\sigma$ level. In blue, the \texttt{Xillver} model and in red, the \texttt{XillverCp} model.
    \texttt{XillverCp} shows higher values of $\Gamma$ (i.e., 1.88-1.92) than \texttt{Xillver}.} 
     \label{fig:xilver_xilvercp_cont}
    \end{figure}
 
 \begin{figure}
    \resizebox{\hsize}{!}{\includegraphics{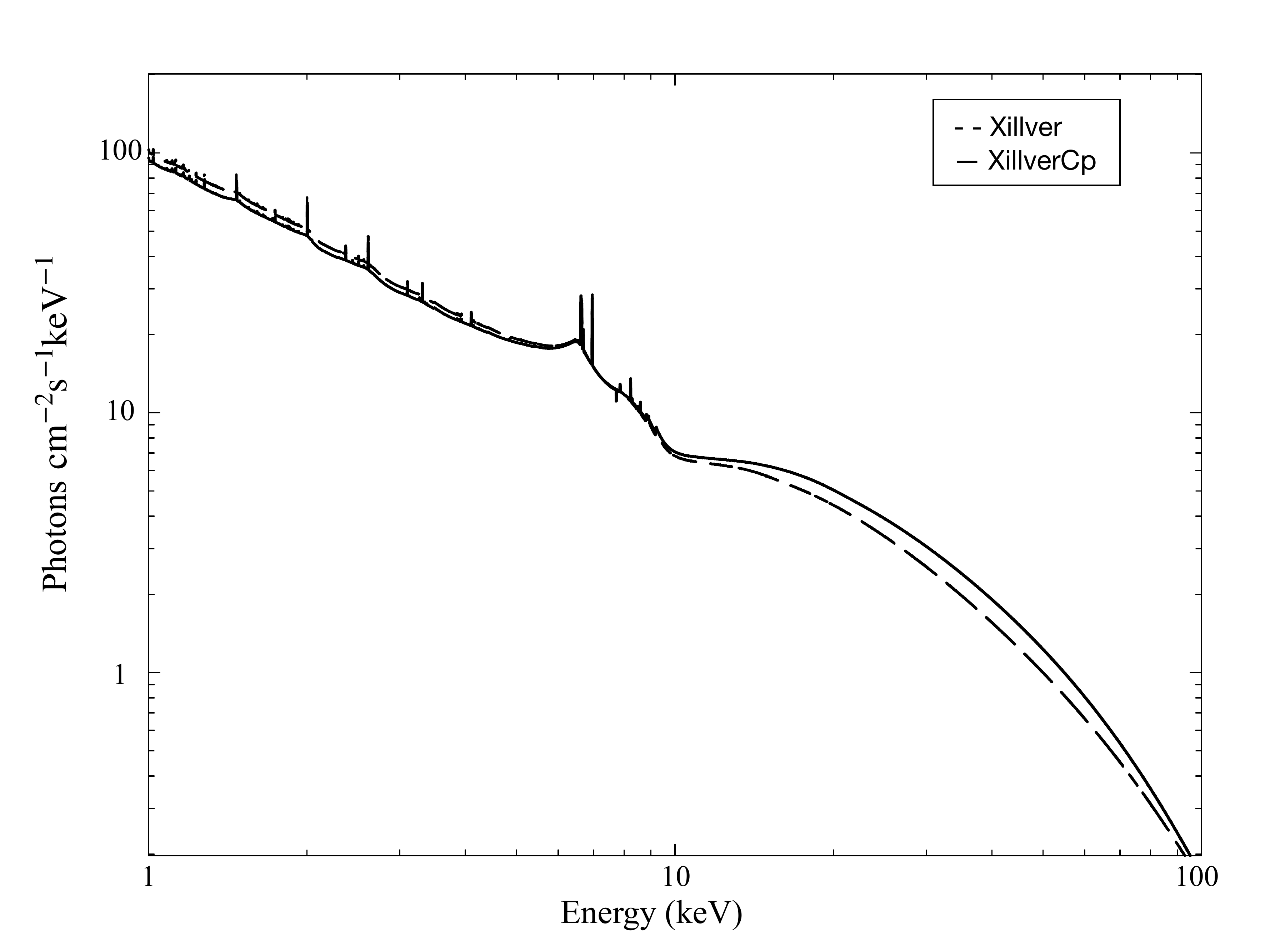}}
    \caption{Comparison of the reflected spectra as calculated with \texttt{Xillver} (dashed line) and \texttt{XillverCp} (line), for an illumination with $\Gamma=$1.9, energy cut-off 100 keV, inclination 30 degrees and the ionization parameter of the disc $\log (\xi)$=3.5, similar to the best-fitting parameters for our data. Note that the small difference in the high energy slope causes the difference in best-fitting cut-off energy of the power-law, which is constrained to be below 160 keV in the case of \texttt{Xillver} and below 250 keV in the case of \texttt{XillverCp}.  }
   \label{fig:xillver_vs_xillvercp}
 \end{figure}

    \begin{figure}
    \resizebox{\hsize}{!}{\includegraphics{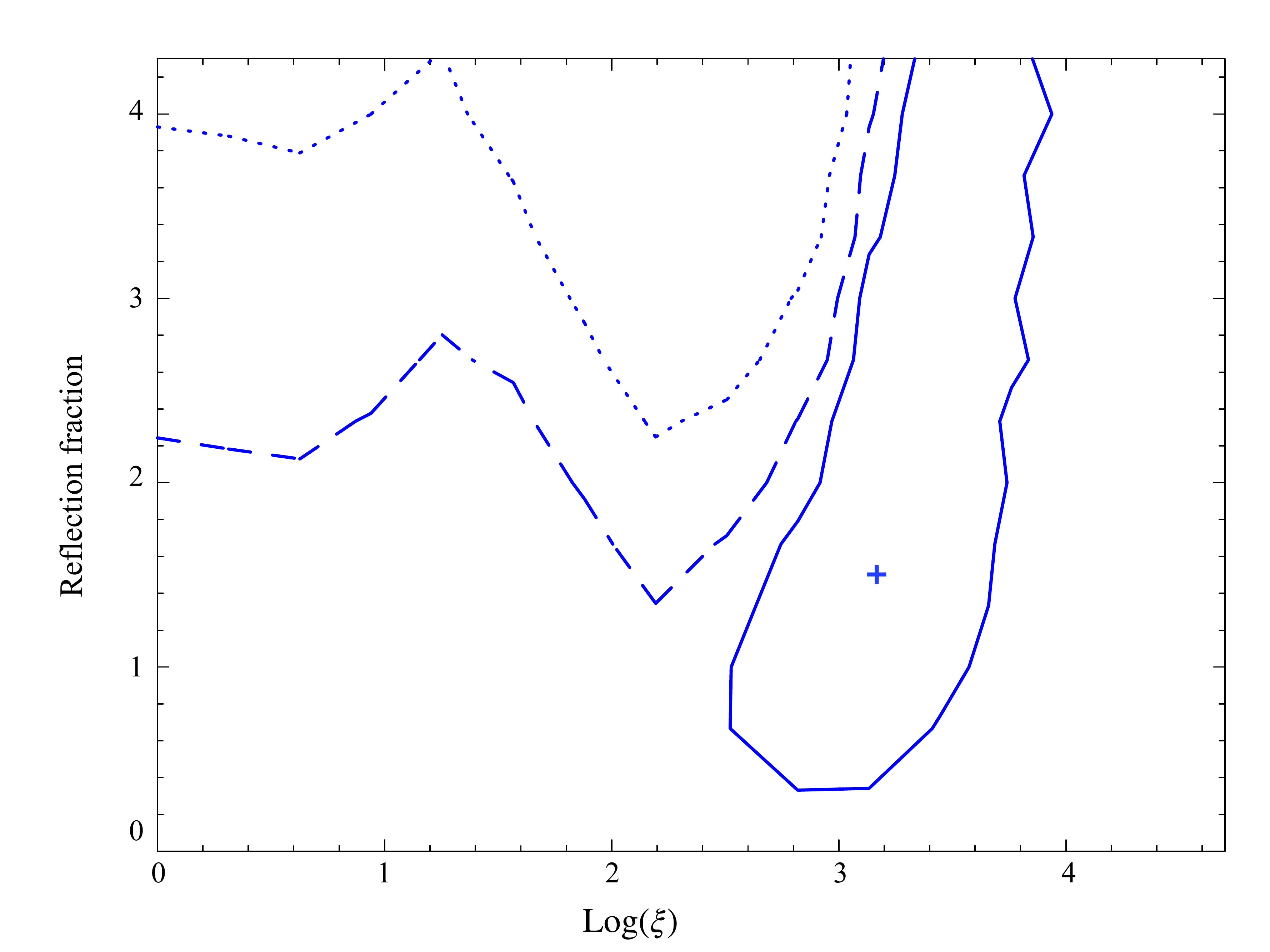}}
    \caption{Two-dimensional $\Delta \chi^{2}$ contours for reflection fraction  $R_{f}$ and disc ionization $\log (\xi)$ calculated with \texttt{Xillver} model. The reflection fraction has as lower limit of $R_{f} >0.3$ and the upper limit is unconstrained. For this high ionization region, the reflected spectrum is very similar to the incident spectrum, so the reflection fraction and overall normalization are degenerate.}
     \label{fig:xillver_logxi_R} 
    \end{figure}
 
\begin{figure}
    \resizebox{\hsize}{!}{\includegraphics{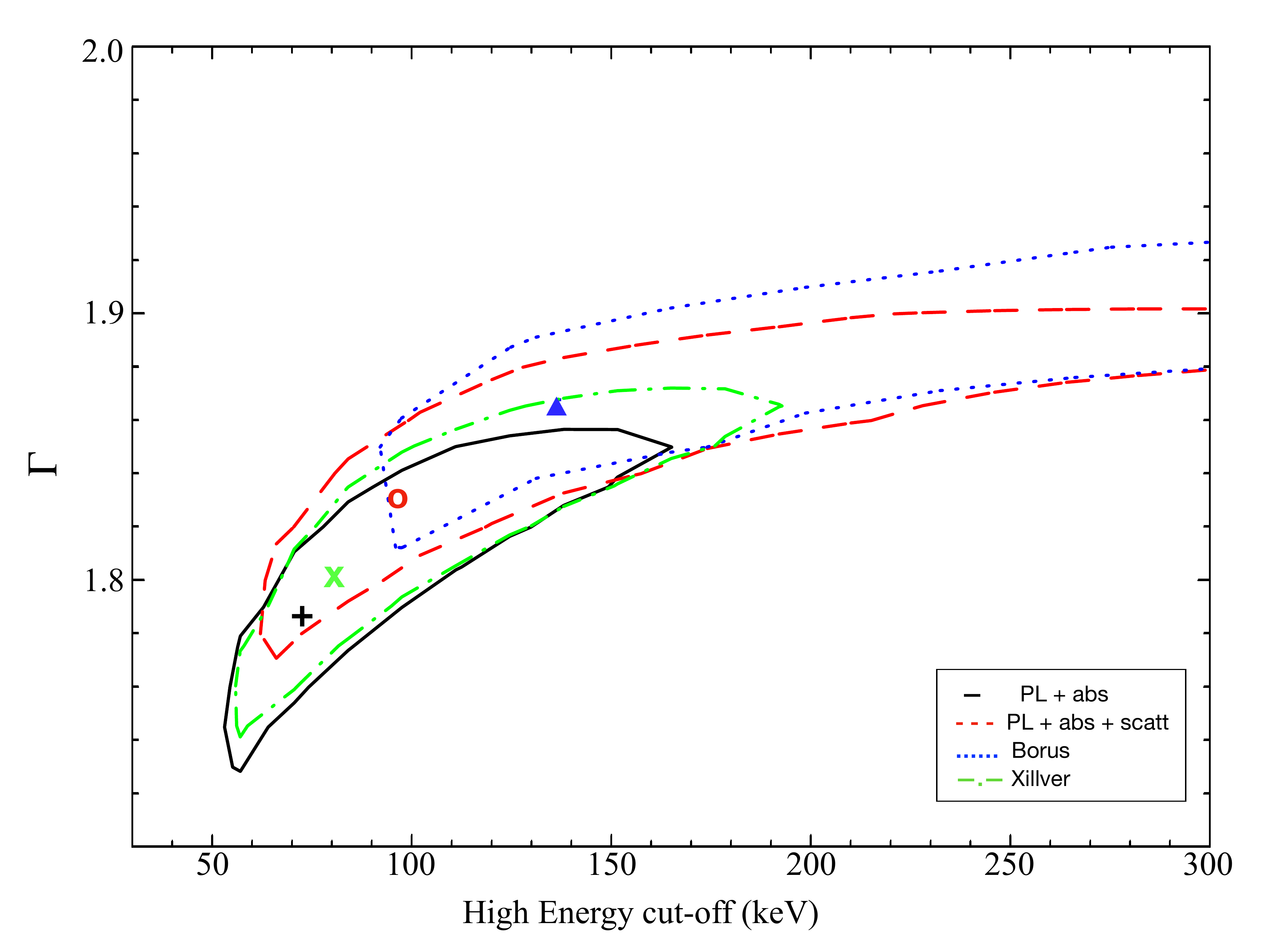}}
    \caption{Two-dimensional $\Delta \chi^{2}$ contours for $\Gamma$ and cut off energies for NGC\,3718. The black line represent a power-law absorbed, in red the Power-law absorbed with an scattered component, in green \texttt{Xillver} reflection model and in blue \texttt{Borus} model. The simplest model shows a $\Gamma$ (i.e., 1.72--1.85) with a low  $E_{\rm cut}$. Adding the scattered component pushes these values slightly higher. Adding the reflection models yields the highest values in the case of torus-like reflector and lower values for a disc reflector. }
     \label{fig:comparacion}
    \end{figure}

 Figure \ref{fig:xilver_xilvercp_cont} shows the allowed ranges of the coronal parameters, $\Gamma$ and the high energy cut-off of the corresponding electron temperature. \texttt{Xillver} uses as model parameter the high energy cut-off of the incident power-law spectrum, while \texttt{XillverCp} uses the electron temperature of the Comptonizing plasma ($kT_{e}$), which is not directly comparable. The electron temperature produces an effective rollover in the Comptonized spectrum at 2 to 3 times higher energy \citep[see, e.g.,][]{petrucci2001}.  Therefore, in the contour plots in Figure \ref{fig:xilver_xilvercp_cont} use different ranges for  $E_{\rm cut}$ of \texttt{Xillver} (top panel) and the electron temperature of \texttt{XillverCp} (bottom panel), considering a conversion factor of 2.5 between them.  With this caveat in mind, there is a good overall agreement between the two models in the best-fitting roll-over energies of the incident spectrum. Nevertheless, \texttt{Xillver} puts a lower upper-limit for $\Gamma$ ($\sim$1.87) compared to \texttt{XillverCp} ($\sim$1.89) at the 1$\sigma$ level and a lower lower-limit with $\sim$1.74 for \texttt{Xillver} and $\sim$1.85 for \texttt{XillverCp}.  

 To illustrate in more detail the difference in the spectral shape of the ionized reflector to the spectrum, we show in Figure \ref{fig:xillver_vs_xillvercp} reflection-only models ($R_{f}=-1$) with one particular value of the $\Gamma$ (1.9),  $E_{\rm cut}$= 100 keV for \texttt{Xillver} and $kT_{e}=40$ keV for \texttt{XillverCp} and inclination of 30 degrees. In this case we fix the ionization parameter to 3.5, similar to the best fitting parameter to our data. The spectral shape is very similar between them; the main difference lies in the high energy tails due to the different shapes of the primary continuum. This small difference in the high energy spectrum can account for the different constraints we found for \texttt{Xillver} and \texttt{XillverCp}, for  $E_{\rm cut}$ and $\Gamma$. The other feature highlighted by this plot is the strong contribution from Compton scattering by the disc, which adds a power-law component to the reflected spectrum.  This highly-ionized disc, therefore, can be consistent with the low amplitude of the reflection features in our data, even if the reflection fraction is high, i.e., without requiring a truncated disc. We further check whether the degeneracy between ionization parameters and reflection fraction can be constrained. Figure \ref{fig:xillver_logxi_R} shows the best fit and contours of allowed parameter values for the reflection fraction $R_{f}$ and $\log\xi$. The 1$\sigma$ contours are restricted to a solution with ionization between 2.5-3.5 with the best value at $\log \xi \sim3.2$ (the case of a highly ionized accretion disc) and reflection fraction $R_f$= 1.6, i.e., more than a half of the sky would be covered by the reflector.   Therefore, the fit with \texttt{Xillver} explains the small reflection features better as an ionized reflector than as a truncated disc.  The reflection fraction has a lower limit of $R_{\rm f}=0.3$ and the upper limit is unconstrained. For this high ionization region, the reflected spectrum is very similar to the incident spectrum, so the reflection fraction and overall normalization are degenerate.

\subsubsection{Comparison between models}

 Our modelling shows two possible configurations to explain the low-amplitude reflection features: a Compton thin reflector covering a high fraction of the sky or a highly ionized accretion disc. From a statistical point of view, these four reflection models, \texttt{MYTorus}, \texttt{Borus02}, \texttt{Xillver} and \texttt{XillverCp}, are  indistinguishable, with only very small differences in the goodness of the fit. From the ratio plots, it appears that the four models perform equally well in describing the data.  Note that there is a general excellent agreement between the values for the column density of the nuclear absorber between the models. 

In Table \ref{tab:resumen} we report the best-fit  $N_{{\rm H},H}$, $\Gamma$ and  $E_{\rm cut}$ values for our source using the \texttt{MYTorus}, \texttt{Borus02}, \texttt{Xillver} and \texttt{XillverCp} models, while in Figure \ref{fig:comparacion} we show the 1$\sigma$ confidence area of the parameter space of the primary emission obtained using the different reflection models.
We found a variation of the best-fitting $\Gamma$ of 0.08 depending on the model. If we use the simple model (an absorbed power-law) we get the lowest value of $\Gamma$ and  $E_{\rm cut}$ (i.e., $\sim$ 76 keV). In case of a power-law with an additional scattered component in the soft band, the values of $\Gamma$ and  $E_{\rm cut}$ increase and show broader extension in the parameter space, indeed, the contour plot at the 1$\sigma$ level is open for  $E_{\rm cut}$. With the inclusion of a disc or torus reflector, we find that the torus reflection model (we plot only the \texttt{Borus} model for visualization purposes) shows higher values in $\Gamma$ and  $E_{\rm cut}$, with only a lower limit for  $E_{\rm cut}$. The ionized reflection model (\texttt{Xillver}) shows lower values for  $E_{\rm cut}$ and $\Gamma$. They  differ, however, in the best-fitting  $E_{\rm cut}$ (81 keV for \texttt{Xillver}, 150 keV for torus reflectors and 30 $kT_{e}$ equivalent to a roll-over energy of $\sim 75$keV for \texttt{XillverCp}) and the allowed range is constrained to be below 200 keV by \texttt{Xillver} while it is unconstrained for all the other reflection models.  

According to our results, $\chi^{2}$ decreases (with $\Delta \chi^2$/d.o.f.=0.02) and the coronal parameters (i.e., $\Gamma$ and  $E_{\rm cut}$) decrease when the reflection component is associated to a disc or increase in the case of a torus. From the statistics point of view, the reflection component is not required because it is not statistically significant.  However, there must be material around the accreting black hole that will produce the weak reflection observed in this galaxy, such as the accretion flow itself and a small BLR \citep{2018cazzoli}. We show that if we do not consider it in the model, the values we obtain from the power-law parameters are overly constrained. For this reason, it is crucial to include a reflection component in the X-ray spectral modeling to estimate accurately the parameters of the primary emission.

 \begin{table*}
\caption{Final compilation of the best-fit models for NGC\,3718 with \emph{XMM-Newton-NuSTAR-Swift/}BAT data.}
\label{tab:resumen}
\begin{tabular}{l|cccccccccc}
\hline
Parameter & \texttt{Power-law} & \texttt{Power-law+scatt} & \texttt{Pexmon} & \texttt{Borus} & \texttt{MYTorus}  & \texttt{Xillver} & \texttt{XillverCp} \\
 (1) & (2) & (3) & (4) & (5) & (6) & (7) & (8)\\
\hline
 $\Gamma$ & 1.79$\pm$0.08 & 1.83$^{+0.08}_{-0.04}$ & 1.84$\pm$0.08& 1.87$^{+0.07}_{-0.08}$ & 1.88$^{+0.06}_{-0.10}$ & 1.82$\pm$0.02 & 1.87$\pm$0.02 \\\\

 $E_{\rm cut}/kT_e$ (keV) & 76$^{+95}_{-22}$ & 97$^{+97}_{-48}$ & 83$^{+86}_{-38}$ & 147$^{+147}_{-83}$ & 150$^{+150}_{-107}$ & 81$^{+137}_{-31}$ & 30$^{+43}_{-12}$\\\\

 $N_{{\rm H},H}$ (10$^{22}$ cm$^{-2})$ & 0.89$^{+0.07}_{-0.06}$ & 1.02$\pm {0.08}$& 1.03$\pm$0.08 & 1.04$^{+0.08}_{-0.09}$ & 1.05$^{+0.06}_{-0.10}$ & 0.98$\pm$0.04 & 1.02$\pm$0.04 \\\\

$C_{\rm {A/B}}$  & 0.99$\pm$0.03 & 0.99$\pm$0.03 & 0.99$\pm$0.03& 0.99$\pm$0.03 & 0.99$\pm$0.03 & 1.00$\pm$0.03 & 1.00$\pm$0.03 \\\\

$C_{XMM}$ & 0.97$\pm$0.05 & 0.98$^{+0.04}_{-0.05}$  & 0.98$^{+0.04}_{-0.05}$ & 0.98$^{+0.04}_{-0.03}$ & 0.98$^{+0.05}_{-0.04}$ &  0.97$^{+0.03}_{-0.02}$ & 0.98$^{+0.03}_{-0.06}$ \\\\

$C_{BAT}$ & 4.7$^{+1.0}_{-0.9}$ & 4.6$\pm 0.9$ & 4.6$^{+0.5}_{-0.9}$ & 4.5$\pm 0.9$ & 4.5$\pm 0.9$  &  4.5$^{+1.0}_{-0.9}$ & 4.4$^{+1.0}_{-0.9}$\\\\

$F^{\rm obs}_{2.0-10.0}$ $_{\rm keV}$ (10$^{-12}$ erg cm$^{-2}$ s$^{-1}$) & 1.39$^{+0.03}_{-0.11}$ & 1.41$^{+0.02}_{-0.14}$ & 1.40$^{+0.01}_{-0.60}$ & 1.41$^{+0.01}_{-0.25}$  & 1.41$^{+0.01}_{-0.16}$ & 1.40$^{+0.05}_{-0.24}$ & 1.39$^{+0.09}_{-0.15}$ \\\\

$F^{\rm int}_{2.0-10.0}$ $_{\rm keV}$ (10$^{-12}$ erg cm$^{-2}$ s$^{-1}$) & 1.52$^{+0.05}_{-0.14}$ & 1.54$^{+0.30}_{-0.32}$ & 1.54$^{+0.26}_{-0.45}$ & 1.55$\pm$0.47  & 1.54$^{+0.62}_{-0.70}$ & 1.53$^{+0.70}_{-0.13}$ & 1.52$^{+0.53}_{-0.35}$ \\\\

$\log(L^{\rm int}_{2.0-10.0}$ $_{\rm keV})$ & 40.57$^{+0.01}_{-0.03}$ & 40.57$^{+0.08}_{-0.10}$ & 40.57$^{+0.07}_{-0.29}$ & 40.57$^{+0.17}_{-0.26}$  & 40.57$^{+0.16}_{-0.26}$ & 40.57$^{+0.60}_{-0.27}$ & 40.57$^{+0.81}_{-0.35}$ \\\\

$F^{\rm obs}_{2.0-70.0}$ $_{\rm keV}$ (10$^{-12}$ erg cm$^{-2}$ s$^{-1}$) & 3.38$^{+0.10}_{-0.07}$ & 3.42$^{+0.07}_{-0.15}$ & 3.40$^{+0.03}_{-0.04}$ & 3.47$^{+0.01}_{-0.14}$  & 3.46$^{+0.16}_{-0.01}$ & 3.46$^{+0.08}_{-0.28}$ & 3.44$^{+0.17}_{-0.22}$ \\\\

$F^{\rm int}_{2.0-70.0}$ $_{\rm keV}$ (10$^{-12}$ erg cm$^{-2}$ s$^{-1}$) & 3.51$\pm$0.05 & 3.57$\pm$0.05 & 3.55$^{+0.06}_{-0.20}$ & 
3.62$\pm$0.80  & 3.61$\pm$0.82 & 3.60$^{+0.63}_{-0.22}$ & 3.68$^{+0.35}_{-0.16}$ \\\\

$\log(L^{\rm int}_{2.0-70.0}$ $_{\rm keV})$ & 40.93$^{+0.06}_{-0.07}$  & 40.93$\pm$ 0.06  & 40.93$^{+0.17}_{-0.28}$ & 40.94$^{+0.09}_{-0.11}$ & 40.94$^{+0.12}_{-0.11}$ & 40.94$^{+0.58}_{-0.26}$ & 40.94$^{+0.60}_{-0.25}$ \\\\

\hline
$\chi^{2}$/d.o.f & 418.96/367 & 409.93/366 & 408.98/365 & 405.56/364 & 406.58/364 & 409.59/363 & 407.21/363\\
\hline
$\chi^{2}_{\nu}$ & 1.142 & 1.120 & 1.121 & 1.114 & 1.117 & 1.128 & 1.122 \\
\hline
\end{tabular}
   \begin{minipage}{17cm}
    \textbf{Note:} Parameter for every model employed to fit the 0.7-79.0 keV data (Col. 1).  $E_{\rm cut}/kT_{e}$ (keV) represents the high energy cut off for \texttt{Xillver} model and the temperature of the gas for \texttt{XillverCp},  $N_{{\rm H},H}$ is the hydrogen column density for the hard component or nuclear. Simple model: power-law+absorption (Col. 2), power-law+absorption+scattered component (Col. 3), \texttt{Pexmon} model (Col. 4), \texttt{Borus} model (Col. 5), \texttt{MYTorus} model (Col. 6),  \texttt{Xillver} model (Col. 7), \texttt{XillverCp} model (Col. 8). $F^{\rm obs}$ and $F^{\rm int}$ correspond to observed and obscuration-corrected fluxes, respectively. $L^{\rm int}$ represents the intrinsic luminosity. We used a distance of D=13Mpc, within the range of Tully-Fischer distance estimates (https://ned.ipac.caltech.edu/). $C_{A/B}$ represents the cross normalization constant between FPMA and FPMB for \emph{NuSTAR}, $C_{XMM}$ for \emph{XMM-Newton} and $C_{BAT}$ for \emph{Swift/}BAT. All the errors correspond to 90$\%$ confidence. $\chi^{2}_{\nu}$ represents $\chi^{2}$ reduced.
    \end{minipage}
\end{table*}

\subsubsection{Comparison with previous results}

We calculate the intrinsic luminosity to compare with the luminosity measured 12 years earlier with \emph{XMM-Newton}. We found a luminosity 8\% lower compared with that previously found by \cite{2014Lore} fitting two absorbed power-laws and 33\% lower compared with \cite{2011younes} using an absorbed power-law. Also, we compare the values obtained by these authors for the $\Gamma$,  $N_{{\rm H},S}$ and  $N_{{\rm H},H}$. \cite{2011younes} fitted an absorbed power-law model to the X-ray spectrum below 10 keV.  They found a $\Gamma$=1.7$\pm$0.1, highly consistent with the one obtained in case of a simple power-law (model in black in Figure \ref{fig:comparacion}), but low compared to the fits including reflection components. \cite{2014Lore} estimated $\Gamma$ by fitting a two power-law model with different absorbing column densities and found $\Gamma$=1.79$^{+0.13}_{-0.08}$. Their result is consistent with ours when fitting a simple power-law. Furthermore, they found consistent values of the hydrogen column density for the nuclear component. In either case, the inconsistency between the parameters previously obtained and our results can be attributed to the use of simpler models over a smaller energy range than the analysis presented here. We recall that this is the first time that high quality data at energies above 10 keV are presented for this source, allowing better constraints on the spectral parameters, including reflection.

Likewise, \cite{2017ricciapJS} using broad-band X-ray spectroscopy (0.3--150 keV) combining \emph{XMM-Newton}, \emph{Swift}/XRT, \emph{ASCA}, \emph{Chandra}, and \emph{Suzaku} observations with \emph{Swift}/BAT data and found a value of $\Gamma$ consistent with our results for the absorbed power-law model. They were not able to constrain  $E_{\rm cut}$ from those data.

Finally, \cite{2018Ricci} found from \emph{Swift}/BAT spectra in the 14--195 keV energy range of a large sample of AGN that  $E_{\rm cut}$ is inversely proportional to the Eddington ratio: sources with Eddington ratio $R_{\rm Edd}$ < 0.1 tend to have  $E_{\rm cut}$ of about 370 keV, while ones with  $R_{\rm Edd}$ > 0.1 possess lower cut-off energies, with  $E_{\rm cut}$ $\sim$160 keV. In the case of NGC 3718 the value of this parameter obtained in all models represents an outlier in this correlation (according to the low Eddington ratio of NGC\,3718) but a high value of the cut-off energy is still consistent within $1\sigma$ level with all reflection models except \texttt{Xillver}.

 \section{Discussion}
 \label{sec:4}
 
In this work we report for the first time the analysis of the broad-band, 0.5--110 keV, emission from the low-luminosity AGN (LLAGN) NGC\,3718, observed simultaneously with \emph{NuSTAR}, \emph{XMM-Newton}, together with archival \emph{Swift}/BAT data. In the following, we discuss the main results of this work.

\subsection{Variability}

The \emph{NuSTAR} observations were taken over 10 days for stretches of 25--90 ks, totalling 230 ks. This allowed us to look for variability on day timescales. We do not detect variability, with an upper limit of $\sigma^{2}_{NXS}<0.08$. This result conforms to the typical behaviour that most LLAGN do not show variability on day timescales, even in the X-ray range \citep{2009Binder, 2011younes, 2018young}. According to \cite{2006mchardy}, who used a small sample of AGN for the study, the characteristic variability timescale of AGN is related to the black hole mass and accretion rate. 
This relation was also explored by \cite{2012omaira} using 104 AGN and later updated by \cite{2018omaira} taking into account absorption effects.
Using the mass of NGC\,3718 ($\log(M_{BH})$=7.85 given by \citealt{2014Lore} and its accretion rate in terms of the Eddington rate ($R_{Edd}  \sim 1.1\times 10 ^{-5}$) the largest amplitude variations are expected on timescales of 
several years using the variability plane reported by these authors.
The analysis of other X-ray data sets have shown variations on timescales of years for NGC\,3718 \citep{2014Lore}, in agreement with the expectation from the variability plane in \cite{2006mchardy}. 
 
\ngc\ was observed once with \emph{Chandra} in 2003 and twice with \emph{XMM-Newton} in 2004. \cite{2011younes} and \cite{2014Lore} studied these data and reported a variable flux in the 2--10 keV energy band by 55\% and 35\%, respectively, on a timescale of one year. They studied also short-term variability from the analysis of the light curves and found no changes on day timescales. This is in agreement with our analysis, where variations on a timescale of ten days were not detected. The variations found by \cite{2011younes} and \cite{2014Lore}, however, can explain the differences in normalization between the \emph{NuSTAR} and \emph{Swift}/BAT data, since the \emph{Swift}/BAT spectrum is the median of data taken over 70 months between December 2004 and September 2010, whereas the \emph{NuSTAR} data were taken outside this period and several years later, in 2017.
 
\subsection{Reflection}

An important feature in the spectra of AGN is the reflection that imprints its mark at X-ray energies. The shape of this reflection component is usually characterized by the FeK$\alpha$ emission line and the Compton hump peaking at $\sim$ 30 keV \citep{1990Pounds, 1994Nandra}. As can be seen in Figure 2, the spectrum of NGC\,3718 shows a weak FeK$\alpha$ line and Compton hump, suggesting a low reflection fraction, as was confirmed with the \texttt{pexmon} reflection model obtaining  $R_{f}$<0.67, with the best fitting value in $R_{f}\sim 0.3$ (30\%) and a $R_{f}$=0 contained within the 1-$\sigma$ contours for E$_{\rm cut}$ between 100 and 250 keV. 
 This is in agreement with other studies of LINERs where the reflection fraction is small (\citealt{2019younes}  and \citealt{2019natalia}, with 5\% and 10\% respectively).
Furthermore, the physical structure causing the reflected spectrum is under debate and different possibilities have been proposed to explain its origin. On the one hand, distant absorbing material such as the torus or clouds in the BLR have been proposed as responsible for the observed reflected emission \citep{2011brightmanNandra}, or even gas in the host galaxy further away from the nucleus and unrelated to the AGN \citep{2014arevalo,2015bauer},
while other authors have argued in favour of the accretion disc as responsible for the reflection \citep{2006Fabian}. In fact, the most plausible scenario is that reflection originates from a combination of all three structures. 

In the case of NGC 3718, we find that the reflection is weak (R$_{f}<0.67$ at $1\sigma$, R$_{f}<0.88$ at $2\sigma$ level) but with a best-fitting value of R$_f=0.3$, showing that the inclusion of reflection provides a better description of the data although, possibly due to the limited counts, the improvement is not statistically significant. The presence of a weak iron line and R$_f=0.3$ is consistent with the fact that the accreting black hole cannot be completely isolated, there must be material around it producing some reflection. Consequently, this component should not be ignored, since leaving it out of the model  could lead to overly constrained values in the coronal parameters. In an effort to characterize the properties of the reflector in the LLAGN NGC\,3718, we have used different X-ray reflection models. All of the models provide equally good fits to the data, but it is worth noting the differences among the reflection component in order to determine the parameters of the power-law ($\Gamma$ and  $E_{\rm cut}$) as well as their physical implications. 

In the case of \ngc\ we detect an absorbing column density of $N_{{\rm H},S}\sim10^{21}$ cm$^{-2}$ on the soft X-ray emission, which we ascribe to a physically extended component, while the nuclear power-law component is under a column density of $N_{{\rm H},H}\sim10^{22}$  cm$^{-2}$. \ngc\ has a prominent dust lane which runs across the entire stellar bulge, and a warped molecular and atomic gas disc, with column density between $\sim$10$^{19-20}$ cm$^{-2}$ \citep{2005krips, 2009sparke}. The average column density of this gaseous disc, however, is too low to explain the obscuration, accounting for at most 10\% of the value measured in the soft X-ray spectrum, and up to 1\% of the column density detected on the nuclear component. From this comparison we can conclude that the absorption we detect is likely related to the active nuclear structure, such as the BLR, the torus, or the narrow line region (NLR). 

On the assumption that all the reflection is produced by distant clouds like the torus, we found that a reflector modelled with either \texttt{MYTorus} or \texttt{Borus02} should be Compton thin and potentially cover a large fraction of the sky although smaller covering fractions (down to 0.2) are also possible within $1\sigma$ of confidence level. A relation between the covering factor of Compton thin material and the accretion rate was previously reported by \cite{2017Riccii}.  They show that accreting black holes with $R_{\rm Edd}$=10$^{-5}$ (the case of NGC 3718) have Compton thin obscurers with covering factors between 0.2-0.6, consistent with our results (see the contour plot in Figure \ref{fig:BORUS_cont1}).

A disc-wind scenario was proposed as an explanation for the torus evolution in LLAGN by \citet{2006elitzur}. This approach establishes that the accretion disc emits vertical winds. For the region inside the dust sublimation radius ($R_{d}$), the gas has no dust, is ionized and forms the BLR, while the wind outside $R_{d}$ has dust and forms the torus.  According with this model, in sources with low accretion rates and low luminosity, the radial column density of the wind is too low  ($N_{H} < 10^{22}$ cm$^{-2}$) to produce detectable emission lines, so the BLR disappears. As the torus is generated by the same mechanism, it would also have a lower column density for lower luminosity, and accretion rate objects.

Later, \cite{2009elitzur}  constructed the distribution of Eddington ratio vs black hole masses and vs bolometric luminosity for objects separated by spectral classification.
They used a sample of AGN from the Palomar spectroscopic survey \citep{1997Ho} with measurements of black hole mass and X-ray luminosity (2--10 keV) available in the literature. 
They show that under a division (corresponding to the theoretical prediction of the disc-wind scenario) in accretion rate and luminosity there are only type 2 objects (Figure 1 in their work) supporting the disappearance of the BLR below the threshold. Also, with these data they estimated the missing constant in their theoretical model to set the limit from where the BLR is observable. According to its accretion rate and luminosity, NGC 3718 falls above the threshold. The BLR and torus, with low column density, are therefore expected, since it is in the region where both AGN types, 1 and 2, are observed. According to the disc wind scenario, \ngc\ should therefore also have a torus, although possibly of low column density.
This galaxy is optically classified as a LINER 1.9 \citep{1997Ho}, which means by definition that only the broad H$\alpha$ is detectable \citep{1989osterbrock}.
The broad H$\alpha$ component in general can either arise from the BLR or from an outflow but  \cite{2018cazzoli} showed that in \ngc\ the broad line is from the BLR. Our observation of the gas column density of the reflector in the X-ray spectrum is $N_{H}\sim6 \times 10^{22}$ cm$^{-2}$. This value shows that the total amount of gas around the AGN in \ngc , combining the BLR and torus, is close to the threshold column density where the BLR is no longer observable \citep{1990netzer}. Therefore the weakness of the optical broad emission lines and of the reflection features together point to a small amount of total gas in the vicinity of the black hole, whether dusty or not.

The disappearance of the dusty torus (e.g., the dusty section of the wind in the disc wind model) at low luminosity is also demonstrated through the evolution of the dust \emph{emission} in the IR. \cite{2017Gonzalez-martin}, using mid-infrared (MIR) spectra from \emph{Spitzer}/IRS of AGN with bolometric luminosities ranging over more than six orders of magnitude, separated the torus emission from other components in the spectra. They reported a gradual reduction in the contribution of the torus with luminosity, with no presence of the torus below $\log[L_{\rm Bol}$ (erg s$^{-1})] <$ 41. For $\log[L_{\rm Bol}$ (erg s$^{-1})] >$ 43, they found that the torus contribution to the bolometric luminosity  has to be larger than 40\%, while for $\log[L_{\rm Bol}$ (erg s$^{-1})] <$ 42 the contribution is less than 20\%. This result is also compatible with the \cite{2006elitzur} model, which indicates less material in the wind for lower luminosities and accretion rates.
According to the luminosity of NGC\,3718, it should fall in the second category of the \cite{2017Gonzalez-martin} work, i.e., showing the presence of absorbing material around the SMBH (the BLR or the torus) with a different configuration compared to more powerful AGN, although a MIR spectrum of this galaxy is not available. 

On the other hand, if the reflected spectrum is dominated by emission from the accretion disc, our data shows that it has to be highly ionized. Key features observed in more powerful AGN are the broad FeK$\alpha$ emission line that can be related to reflection from the accretion disc \citep{2009fabian, brenneman2011spin, ricci2014suzaku} or a correlation between the ionization parameter with the Eddington ratio \citep{ballantyne2011, keek2016}. However, similar studies have not been possible for LINERs given the weakness of the FeK$\alpha$ emission line, as the case for NGC\,3718. Moreover, X-ray reflection models of highly ionized discs have not been performed for other LINERs, preventing us from any comparison with other works. It is worth noting that the geometry of the inner parts of LINERs might differ from more powerful AGN, implying, for instance, that the disc could be truncated (\citealt{2009Gu,2011younes, 2013Lore, 2016Lore, 2018She}), although our fits with both \texttt{Xillver} and \texttt{XillverCp} prefer an ionized disc to a truncated one to explain the small amplitude of the reflection features (see Figure\ref{fig:xillver_logxi_R}).

Therefore, we propose that \ngc\ has a torus/BLR that contributes at least partially to the reflection spectrum, whereas our study prevents us from confirming the presence of reflection off the accretion disc. 

\subsection{Accretion mechanism}
 The coronal emission of NGC\,3718 is fitted by an absorbed power-law with $\Gamma \sim$ 1.8 and $N_{{\rm H},H} \sim $ 10$^{22}$ cm$^{-2}$. This $\Gamma$ is consistent with typical measurements for AGN, suggesting the same physical origin for the X-ray emission \citep{2011brightmanNandra}.
 
Spectral differences between high and low-luminosity AGN may arise from their accretion mechanism. While the standard accretion disc explains well the powering of highly accreting AGN \citep{1973shakura}, it has been suggested that for LLAGN the emission mechanism becomes inefficient and the X-ray emission originates in ADAFs \citep{2009Gu,2011younes, 2013Lore, 2016Lore, 2018She} similar to that in X-ray binaries (XRBs) in their low/hard state \citep{2007ma, 2010ueda, 2011xu}. AGN are thought to be scaled up versions of Galactic black hole X-ray binaries. The study of the accretion mechanism in XRBs and AGN has been approached by relating $\Gamma$ to the Eddington ratio ($R_{\rm Edd}$ =  $L_{\rm Bol}$/ $L_{\rm Edd}$). This relation shows a positive trend (soft state in XRBs) for high luminosity sources above a threshold value of $R_{\rm Edd}$ and negative trend (low state for XRBs) below this threshold \citep{2009Gu}. However, in the case of LLAGN, which fall in the anti-correlation section of this relation, it shows a high dispersion that is still not understood --- it could be due to the sensitivity of the measurements or to intrinsic diversity of the nuclei. Estimating $\Gamma$ using high-quality X-ray data and studying how sensitive $\Gamma$ is to the reflection model used in the fit is an important step to constrain the origin of the scatter in this relation. 

Our best-fitting $\Gamma$ for NGC\,3718, including the reflection component, falls on the mean value of the correlation $\Gamma$ vs $R_{\rm Edd}$ given by \cite{2009Gu}, \cite{2011younes}, and \cite{ 2018She}. The coincidence between the measured and expected values of $\Gamma$ suggests that high quality X-ray spectra, together with modelling including the reflection component, can reduce the large scatter seen in this correlation.  

Furthermore, we can study the position of \ngc\ in the fundamental dichotomy between the local radio AGN population (see \citealt{2012best}).  This plane proposes that according to the optical spectra, AGN can be classified as quasar mode/HERG (high-excitation) where the material is accreted onto the black hole through a radiatively-efficient, optically-thick, geometrically thin accretion disc (e.g., \citealt{1973shakura}) or radio mode/LERG (low-excitation) dominated by ADAFs. The excitation level of the emission line region is expected to be defined by the hardness of the UV spectrum from the central source, therefore, a thin disc reaching the innermost orbit would produce a hard ionizing UV continuum and more high excitation lines, while a truncated disc, replaced in the innermost regions by an ADAF, would produce a low-excitation spectrum. The excitation index (EI) defined by \citet{2010buttiglione} is used by \citet{2012best} to separate low from high-excitation sources, with a threshold at EI=0.95. The optical spectrum of \ngc\ characterized by \citet{2006Moustakas} results in EI=0.89, in the low-excitation region but very close to the high-excitation threshold. The equivalent width of [OIII] is alternatively used as an excitation quantifier and \citet{2012best} propose 5\AA\ as the threshold value. With this criterion \ngc\ also falls in the low-excitation region but at the high excitation end of LERGs, with an [OIII] EW=2.6\AA\ \citep{2006Moustakas}. Based on these criteria, NGC 3718 can be classified as a low excitation galaxy that is in favor of an ADAF instead of accretion via a geometrically thin disc, although it is close to the limit between low and high-excitation sources.

\subsection{Source of X-ray emission}

The source of X-ray emission is generally unresolved in LLAGN and its origin is under debate: it might be the ADAF itself or it might be synchrotron emission from a jet. 

The jet origin is supported by the fact that low-luminosity objects tend to be radio loud, as noted by \cite{2002ho}, who shows that radio loudness anti-correlates strongly with $R_{\rm Edd}$.  According with this relation, the accretion rate in NGC\,3718 should result in a radio loud classification. Moreover, \ngc \ was observed with VLBA \citep{2005neil} and MERLIN \citep{2007krips, 2015markakis} and they reported extended emission at 18 cm with signs of a compact (0.5\arcsec or 34 pc) radio jet detected at 4$\sigma$ significance which is weakly present at 6 cm as well. Therefore we need to take into account the possibility that this jet emits X-rays.

Our next step is thus to study the jet dominance in this galaxy. \cite{2012younes} classified a sample of LINER 1s into radio-loud or radio quiet classes according to the radio loudness parameter ($R_{x}$=$\nu L_{\nu}$(5 GHz)/$L_{2-10 \rm keV}$, \citealt{2003terashima}) which compares the radio to X-ray fluxes. In agreement with this quantity, a LLAGN can be classified as radio-loud if $\log(R_{x})$>-4.5. For NGC 3718, \citealt{2012younes} found $\log(R_{x})$=-3.81, classifying this galaxy as a radio-loud. Using the X-ray luminosity from \cite{2011younes}, \cite{2014Lore} and our results, we calculate slightly different values of $R_{x}$, but consistent with a radio loud classification. \citet{2007panessa}, however, challenged this simple criterion to judge radio loudness. These authors studied the Seyferts in the Palomar sample \citep{1997Ho}, which they assumed as radio-quiet, and compared to low-luminosity radio galaxies, i.e., radio-loud, and reported that for low-luminosity AGN the limit for radio-loud should be larger. They showed that a better threshold is $\log(R_{x})\sim$-2.8, placing \ngc\ in the radio-quiet regime. Furthermore, \cite{2007maoz} compared the radio-to-UV fluxes of a sample of AGN covering a wide range in luminosity and found that $R_{UV}=L_{\nu}$(5 GHz)/$L_{\nu}$ (2500\AA) increases with decreasing luminosity for both radio loud and radio quiet populations, so a more natural threshold should be a function of UV luminosity. Taking the values of $R_{UV}$ and $L_{\nu}$(2500 \AA) from \cite{2017Li}, \ngc\ falls in the region between radio-loud and radio-quiet, for its UV luminosity, in the $R_{UV}$ vs $L_{UV}$ plane of \cite{2007maoz}. In agreement with this result, \cite{2012younes} present the SED of NGC\,3718 and compared it with the \cite{1994elvis} average SEDs from a sample of radio-loud and radio-quiet AGN, finding that the radio emission of \ngc\ falls exactly in the middle of both models. Taking all the information together, the radio loudness for \ngc\ remains unclear. Nevertheless, it is worth remarking that under the criteria that take into account luminosity, and also by the shape of the SED, the radio loudness of \ngc\ is at most borderline, so it is unlikely that the radio jet emission would dominate the X-ray spectrum. 

The best way to know if the radio jet dominates in the X-rays would be to confirm or reject the spectral curvature in the nuclear emission,  since the ADAF model predicts a curvature while the synchrotron jet emission would be a pure power-law. Our data, however, do not allow us to confirm or rule out this feature; depending on the reflection model used, a cutoff in the nuclear power-law is required or not. A detailed physical modelling of a more complete SED will hopefully reveal the dominance of one of these physical mechanisms in this LLAGN.

    \section{Summary}
    \label{sec:5}
Through simultaneous \emph{NuSTAR + XMM--Newton} plus archival \emph{Swift}/BAT observations, we performed a variability and spectral analysis of the LLAGN NGC\,3718. The summary of our main results are reported in the following:
\begin{itemize}
    \item We  do  not  detect  any  significant  variability  in the nucleus of  NGC\,3718 within the \emph{NuSTAR} observations, on a timescale of 10 days.

  \item The NGC\,3718 obscuration corrected flux in the 2--10 keV energy band is  8\% lower than the value previously reported by \cite{2014Lore} and 33\% lower than the value reported by \cite{2011younes} using \emph{XMM--Newton} data from 12 years ago. 

    \item The X-ray spectrum shows a small Fe K$\alpha$ line, indicative of a reflection component. A simple fit including neutral reflection with \texttt{pexmon} indicates R$_f$ <0.67 with the best-fitting reflection fraction  $R_{f}$=0.3, although a $R_{f}$=0 is also allowed within the $1-\sigma$ contours. Even though the reflection is weak it should not be ignored as this could lead to a misinterpretation of the coronal parameters as explained in Sec.\ref{sec:4}. 
    
    \item The type of reflector affects the measurement of the power-law parameters. Both the $\Gamma$ and the cut-off energy are marginally lower for a disc reflector than for a torus. While one of the disc reflectors results in a low and bounded cut-off energy, the torus reflectors produce a best fitting cut-off energy above the observed energy range and unconstrained to higher values.  Therefore we cannot confirm or rule out curvature in the continuum in this spectral range. 

    \item We cannot differentiate between the four reflection models fitted to the data, but these fits allow us to put constraints on each physical scenario. Reflection dominated by a smooth, neutral torus, as modelled be \texttt{MYTorus} or \texttt{Borus02} should be Compton thin and preferentially cover a large fraction of the sky, although covering fractions as low as 0.1 are still allowed by the data at 1 $\sigma$ level. In the case of an ionized disc dominating the reflected spectrum, as modelled by \texttt{Relxill}, a highly ionized disc is required.
    
    \item The column density obtained for a neutral reflector, compatible with the small features seen in the X-ray spectrum, is $N_{\rm H}\sim 6\times 10^{22}\rm{cm}^{-2}$. This is similar to the column density observed in absorption $N_{{\rm H},H}\sim  10^{22}\rm{cm}^{-2}$ and of the same order of magnitude of the limiting column density for an observable BLR. 
    
    \item Our results show the importance of including the reflection
    when analyzing the accretion mechanism in LLAGN and to understand degeneracies with the intrinsic power-law parameters.

\end{itemize}
The application of the methodology explained here will be subject of a forthcoming paper using a sample of AGN covering a large range in $R_{\rm Edd}$ to estimate the intrinsic $\Gamma$ and $E_{\rm cut}$ with high accuracy in order to study the accretion mechanism in LLAGN, as well as the physical origin of the reflected spectrum.

More observations are needed to test the paradigm of the reflection in AGN. Future missions like \emph{HEX-P} may allow to extend this work. This next-generation high-energy X-ray observatory will have broad band (0.1-200 keV) and 100 times the sensitivity of any previous mission, allowing the extension of this work for a sample of LLAGN with low accretion rate in the spectral region where the reflection is most important. Noting that the work presented here required 200 ks of \nustar\ observations, it is clear that even this level of accuracy is currently only possible for very local AGN. A much larger collecting area will allow similar high quality spectra to be obtained for low luminosity AGN in a larger volume, permitting for example the refinement of the $\Gamma$ vs $L/L_{\rm Edd}$ at the low accretion rate regime. Future improvements in the observations and the acquisition of high-quality data may allow to extent this work for a large sample with smaller acquisition time as well, to study in detail the properties of the reflectors.

    \section*{Acknowledgments}

We thank the referee for a thorough reading of the manuscript that led to an improved version of the article. Y.D and P.A. acknowledge the financial support from CONICYT PIA ACT172033 and the Max-Planck Society through a Partner Group grant. L.H.G. acknowledges financial support from FONDECYT through grant 3170527. O.G.M acknowledge financial support from the UNAM PAPIIT project IA103118. CR acknowledges Fondecyt 11190831. D.M. acknowledge financial support FROM FAPESP (Funda\c{c}\~ao de Amparo \`a Pesquisa do Estado de S\~ao Paulo), under grant 2011/19824-8 (DMN). 
F.E.B. acknowledges CONICYT grants CATA-Basal AFB-170002 (FEB), FONDECYT Regular 1190818 (FEB) and 1200495 (FEB); and Chile's Ministry of Economy, Development, and Tourism's Millennium Science Initiative through grant IC120009, awarded to The Millennium Institute of Astrophysics, MAS (FEB). This work made use of data from the \emph{NuSTAR} mission, a project led by the California Institute of Technology, managed by the Jet Propulsion Laboratory, and funded by the National Aeronautics and Space Administration. This  work  was based  on  observations  obtained  with \emph{XMM-Newton} provided by the \emph{XMM-Newton} Science Archive (XSA). This research has made use of the NASA/IPAC Extragalactic Database (NED) which is operated by the Jet Propulsion Laboratory, of data obtained from the HighEnergy Astrophysics Science Archive Research Center (HEASARC), provided by NASA Goddard Space Flight Center. Part of this work is based on archival data, software or online services provided by the Space Science Data Center - ASI.

\section*{Data availability}
The datasets underlying this article are available in the domain:
https://heasarc.gsfc.nasa.gov/

    \bibliographystyle{mnras}
    \bibliography{ref}

\begin{thebibliography}{}
\makeatletter
\relax
\def\mn@urlcharsother{\let\do\@makeother \do\$\do\&\do\#\do\^\do\_\do\%\do\~}
\def\mn@doi{\begingroup\mn@urlcharsother \@ifnextchar [ {\mn@doi@}
  {\mn@doi@[]}}
\def\mn@doi@[#1]#2{\def\@tempa{#1}\ifx\@tempa\@empty \href
  {http://dx.doi.org/#2} {doi:#2}\else \href {http://dx.doi.org/#2} {#1}\fi
  \endgroup}
\def\mn@eprint#1#2{\mn@eprint@#1:#2::\@nil}
\def\mn@eprint@arXiv#1{\href {http://arxiv.org/abs/#1} {{\tt arXiv:#1}}}
\def\mn@eprint@dblp#1{\href {http://dblp.uni-trier.de/rec/bibtex/#1.xml}
  {dblp:#1}}
\def\mn@eprint@#1:#2:#3:#4\@nil{\def\@tempa {#1}\def\@tempb {#2}\def\@tempc
  {#3}\ifx \@tempc \@empty \let \@tempc \@tempb \let \@tempb \@tempa \fi \ifx
  \@tempb \@empty \def\@tempb {arXiv}\fi \@ifundefined
  {mn@eprint@\@tempb}{\@tempb:\@tempc}{\expandafter \expandafter \csname
  mn@eprint@\@tempb\endcsname \expandafter{\@tempc}}}

\bibitem[\protect\citeauthoryear{{Antonucci}}{{Antonucci}}{1993}]{1993Antonucci}
{Antonucci} R.,  1993, \mn@doi [\araa] {10.1146/annurev.aa.31.090193.002353},
  \href {http://adsabs.harvard.edu/abs/1993ARA%26A..31..473A} {31, 473}

\bibitem[\protect\citeauthoryear{{Ar{\'e}valo} et~al.,}{{Ar{\'e}valo}
  et~al.}{2014}]{2014arevalo}
{Ar{\'e}valo} P.,  et~al., 2014, \mn@doi [\apj] {10.1088/0004-637X/791/2/81},
  \href {https://ui.adsabs.harvard.edu/abs/2014ApJ...791...81A} {791, 81}

\bibitem[\protect\citeauthoryear{{Arnaud}}{{Arnaud}}{1996}]{1996arnaud}
{Arnaud} K.~A.,  1996, in {Jacoby} G.~H.,  {Barnes} J.,  eds,  Astronomical
  Society of the Pacific Conference Series Vol. 101, Astronomical Data Analysis
  Software and Systems V. p.~17

\bibitem[\protect\citeauthoryear{{Ballantyne}, {McDuffie}  \&
  {Rusin}}{{Ballantyne} et~al.}{2011}]{ballantyne2011}
{Ballantyne} D.~R.,  {McDuffie} J.~R.,   {Rusin} J.~S.,  2011, \mn@doi [\apj]
  {10.1088/0004-637X/734/2/112}, \href
  {https://ui.adsabs.harvard.edu/abs/2011ApJ...734..112B} {734, 112}

\bibitem[\protect\citeauthoryear{{Balokovi{\'c}} et~al.,}{{Balokovi{\'c}}
  et~al.}{2018}]{2018balokovic}
{Balokovi{\'c}} M.,  et~al., 2018, \mn@doi [\apj] {10.3847/1538-4357/aaa7eb},
  \href {http://adsabs.harvard.edu/abs/2018ApJ...854...42B} {854, 42}

\bibitem[\protect\citeauthoryear{{Bauer} et~al.,}{{Bauer}
  et~al.}{2015}]{2015bauer}
{Bauer} F.~E.,  et~al., 2015, \mn@doi [\apj] {10.1088/0004-637X/812/2/116},
  \href {https://ui.adsabs.harvard.edu/abs/2015ApJ...812..116B} {812, 116}

\bibitem[\protect\citeauthoryear{{Baumgartner}, {Tueller}, {Markwardt},
  {Skinner}, {Barthelmy}, {Mushotzky}, {Evans}  \& {Gehrels}}{{Baumgartner}
  et~al.}{2013}]{2013baumgartner}
{Baumgartner} W.~H.,  {Tueller} J.,  {Markwardt} C.~B.,  {Skinner} G.~K.,
  {Barthelmy} S.,  {Mushotzky} R.~F.,  {Evans} P.~A.,   {Gehrels} N.,  2013,
  \mn@doi [The Astrophysical Journal Supplement Series]
  {10.1088/0067-0049/207/2/19}, \href
  {https://ui.adsabs.harvard.edu/\#abs/2013ApJS..207...19B} {207, 19}

\bibitem[\protect\citeauthoryear{{Best} \& {Heckman}}{{Best} \&
  {Heckman}}{2012}]{2012best}
{Best} P.~N.,  {Heckman} T.~M.,  2012, \mn@doi [\mnras]
  {10.1111/j.1365-2966.2012.20414.x}, \href
  {https://ui.adsabs.harvard.edu/abs/2012MNRAS.421.1569B} {421, 1569}

\bibitem[\protect\citeauthoryear{{Binder}, {Markowitz}  \&
  {Rothschild}}{{Binder} et~al.}{2009}]{2009Binder}
{Binder} B.,  {Markowitz} A.,   {Rothschild} R.~E.,  2009, \mn@doi [\apj]
  {10.1088/0004-637X/691/1/431}, \href
  {http://adsabs.harvard.edu/abs/2009ApJ...691..431B} {691, 431}

\bibitem[\protect\citeauthoryear{Brenneman et~al.,}{Brenneman
  et~al.}{2011}]{brenneman2011spin}
Brenneman L.,  et~al., 2011, The Astrophysical Journal, 736, 103

\bibitem[\protect\citeauthoryear{{Brightman} \& {Nandra}}{{Brightman} \&
  {Nandra}}{2011}]{2011brightmanNandra}
{Brightman} M.,  {Nandra} K.,  2011, \mn@doi [\mnras]
  {10.1111/j.1365-2966.2011.18207.x}, \href
  {https://ui.adsabs.harvard.edu/abs/2011MNRAS.413.1206B} {413, 1206}

\bibitem[\protect\citeauthoryear{{Buttiglione}, {Capetti}, {Celotti}, {Axon},
  {Chiaberge}, {Macchetto}  \& {Sparks}}{{Buttiglione}
  et~al.}{2010}]{2010buttiglione}
{Buttiglione} S.,  {Capetti} A.,  {Celotti} A.,  {Axon} D.~J.,  {Chiaberge} M.,
   {Macchetto} F.~D.,   {Sparks} W.~B.,  2010, \mn@doi [\aap]
  {10.1051/0004-6361/200913290}, \href
  {https://ui.adsabs.harvard.edu/abs/2010A&A...509A...6B} {509, A6}

\bibitem[\protect\citeauthoryear{{Cappi} et~al.,}{{Cappi}
  et~al.}{2006}]{cappi2006}
{Cappi} M.,  et~al., 2006, \mn@doi [\aap] {10.1051/0004-6361:20053893}, \href
  {https://ui.adsabs.harvard.edu/abs/2006A&A...446..459C} {446, 459}

\bibitem[\protect\citeauthoryear{{Cazzoli} et~al.,}{{Cazzoli}
  et~al.}{2018}]{2018cazzoli}
{Cazzoli} S.,  et~al., 2018, \mn@doi [\mnras] {10.1093/mnras/sty1811}, \href
  {https://ui.adsabs.harvard.edu/abs/2018MNRAS.480.1106C} {480, 1106}

\bibitem[\protect\citeauthoryear{{Dickey} \& {Lockman}}{{Dickey} \&
  {Lockman}}{1990}]{1990Dickey}
{Dickey} J.~M.,  {Lockman} F.~J.,  1990, \mn@doi [\araa]
  {10.1146/annurev.aa.28.090190.001243}, \href
  {http://adsabs.harvard.edu/abs/1990ARA%26A..28..215D} {28, 215}

\bibitem[\protect\citeauthoryear{{Elitzur}}{{Elitzur}}{2008}]{2008Elitzur}
{Elitzur} M.,  2008, \mn@doi [\nar] {10.1016/j.newar.2008.06.010}, \href
  {http://adsabs.harvard.edu/abs/2008NewAR..52..274E} {52, 274}

\bibitem[\protect\citeauthoryear{{Elitzur} \& {Ho}}{{Elitzur} \&
  {Ho}}{2009}]{2009elitzur}
{Elitzur} M.,  {Ho} L.~C.,  2009, \mn@doi [\apjl]
  {10.1088/0004-637X/701/2/L91}, \href
  {https://ui.adsabs.harvard.edu/abs/2009ApJ...701L..91E} {701, L91}

\bibitem[\protect\citeauthoryear{{Elitzur} \& {Shlosman}}{{Elitzur} \&
  {Shlosman}}{2006}]{2006elitzur}
{Elitzur} M.,  {Shlosman} I.,  2006, \mn@doi [\apjl] {10.1086/508158}, \href
  {https://ui.adsabs.harvard.edu/abs/2006ApJ...648L.101E} {648, L101}

\bibitem[\protect\citeauthoryear{{Elvis} et~al.,}{{Elvis}
  et~al.}{1994}]{1994elvis}
{Elvis} M.,  et~al., 1994, \mn@doi [\apjs] {10.1086/192093}, \href
  {https://ui.adsabs.harvard.edu/abs/1994ApJS...95....1E} {95, 1}

\bibitem[\protect\citeauthoryear{{Fabian}}{{Fabian}}{2006}]{2006Fabian}
{Fabian} A.~C.,  2006, in {Wilson} A.,  ed.,  ESA Special Publication Vol. 604,
  The X-ray Universe 2005. p.~463 (\mn@eprint {} {astro-ph/0511537})

\bibitem[\protect\citeauthoryear{{Fabian} et~al.,}{{Fabian}
  et~al.}{2009}]{2009fabian}
{Fabian} A.~C.,  et~al., 2009, \mn@doi [\nat] {10.1038/nature08007}, \href
  {https://ui.adsabs.harvard.edu/abs/2009Natur.459..540F} {459, 540}

\bibitem[\protect\citeauthoryear{{Garc{\'{\i}}a}, {Dauser}, {Reynolds},
  {Kallman}, {McClintock}, {Wilms}  \& {Eikmann}}{{Garc{\'{\i}}a}
  et~al.}{2013}]{2013garcia}
{Garc{\'{\i}}a} J.,  {Dauser} T.,  {Reynolds} C.~S.,  {Kallman} T.~R.,
  {McClintock} J.~E.,  {Wilms} J.,   {Eikmann} W.,  2013, \mn@doi [\apj]
  {10.1088/0004-637X/768/2/146}, \href
  {http://adsabs.harvard.edu/abs/2013ApJ...768..146G} {768, 146}

\bibitem[\protect\citeauthoryear{{Gonz{\'a}lez-Mart{\'\i}n}}{{Gonz{\'a}lez-Mart{\'\i}n}}{2018}]{2018omaira}
{Gonz{\'a}lez-Mart{\'\i}n} O.,  2018, \mn@doi [\apj]
  {10.3847/1538-4357/aab7ec}, \href
  {https://ui.adsabs.harvard.edu/abs/2018ApJ...858....2G} {858, 2}

\bibitem[\protect\citeauthoryear{{Gonz{\'a}lez-Mart{\'\i}n} \&
  {Vaughan}}{{Gonz{\'a}lez-Mart{\'\i}n} \& {Vaughan}}{2012}]{2012omaira}
{Gonz{\'a}lez-Mart{\'\i}n} O.,  {Vaughan} S.,  2012, \mn@doi [\aap]
  {10.1051/0004-6361/201219008}, \href
  {https://ui.adsabs.harvard.edu/abs/2012A&A...544A..80G} {544, A80}

\bibitem[\protect\citeauthoryear{{Gonz{\'a}lez-Mart{\'{\i}}n}, {Masegosa},
  {M{\'a}rquez}, {Guainazzi}  \&
  {Jim{\'e}nez-Bail{\'o}n}}{{Gonz{\'a}lez-Mart{\'{\i}}n}
  et~al.}{2009}]{2009Gonzalez}
{Gonz{\'a}lez-Mart{\'{\i}}n} O.,  {Masegosa} J.,  {M{\'a}rquez} I.,
  {Guainazzi} M.,   {Jim{\'e}nez-Bail{\'o}n} E.,  2009, \mn@doi [\aap]
  {10.1051/0004-6361/200912288}, \href
  {http://adsabs.harvard.edu/abs/2009A%26A...506.1107G} {506, 1107}

\bibitem[\protect\citeauthoryear{{Gonz{\'a}lez-Mart{\'\i}n}
  et~al.,}{{Gonz{\'a}lez-Mart{\'\i}n} et~al.}{2017}]{2017Gonzalez-martin}
{Gonz{\'a}lez-Mart{\'\i}n} O.,  et~al., 2017, \mn@doi [\apj]
  {10.3847/1538-4357/aa6f16}, \href
  {https://ui.adsabs.harvard.edu/abs/2017ApJ...841...37G} {841, 37}

\bibitem[\protect\citeauthoryear{{Gu} \& {Cao}}{{Gu} \& {Cao}}{2009}]{2009Gu}
{Gu} M.,  {Cao} X.,  2009, \mn@doi [\mnras] {10.1111/j.1365-2966.2009.15277.x},
  \href {http://adsabs.harvard.edu/abs/2009MNRAS.399..349G} {399, 349}

\bibitem[\protect\citeauthoryear{{Haardt} \& {Maraschi}}{{Haardt} \&
  {Maraschi}}{1993}]{1993haardt}
{Haardt} F.,  {Maraschi} L.,  1993, \mn@doi [\apj] {10.1086/173020}, \href
  {http://adsabs.harvard.edu/abs/1993ApJ...413..507H} {413, 507}

\bibitem[\protect\citeauthoryear{{Harrison} et~al.,}{{Harrison}
  et~al.}{2013}]{2013harrison}
{Harrison} F.~A.,  et~al., 2013, \mn@doi [\apj] {10.1088/0004-637X/770/2/103},
  \href {http://adsabs.harvard.edu/abs/2013ApJ...770..103H} {770, 103}

\bibitem[\protect\citeauthoryear{{Heckman}}{{Heckman}}{1980}]{1980Heckman}
{Heckman} T.~M.,  1980, \aap, \href
  {http://adsabs.harvard.edu/abs/1980A%26A....87..152H} {87, 152}

\bibitem[\protect\citeauthoryear{{Hern{\'a}ndez-Garc{\'{\i}}a},
  {Gonz{\'a}lez-Mart{\'{\i}}n}, {M{\'a}rquez}  \&
  {Masegosa}}{{Hern{\'a}ndez-Garc{\'{\i}}a} et~al.}{2013}]{2013Lore}
{Hern{\'a}ndez-Garc{\'{\i}}a} L.,  {Gonz{\'a}lez-Mart{\'{\i}}n} O.,
  {M{\'a}rquez} I.,   {Masegosa} J.,  2013, \mn@doi [\aap]
  {10.1051/0004-6361/201321563}, \href
  {http://adsabs.harvard.edu/abs/2013A%26A...556A..47H} {556, A47}

\bibitem[\protect\citeauthoryear{{Hern{\'a}ndez-Garc{\'{\i}}a},
  {Gonz{\'a}lez-Mart{\'{\i}}n}, {Masegosa}  \&
  {M{\'a}rquez}}{{Hern{\'a}ndez-Garc{\'{\i}}a} et~al.}{2014}]{2014Lore}
{Hern{\'a}ndez-Garc{\'{\i}}a} L.,  {Gonz{\'a}lez-Mart{\'{\i}}n} O.,  {Masegosa}
  J.,   {M{\'a}rquez} I.,  2014, \mn@doi [\aap] {10.1051/0004-6361/201424140},
  \href {http://adsabs.harvard.edu/abs/2014A%26A...569A..26H} {569, A26}

\bibitem[\protect\citeauthoryear{{Hern{\'a}ndez-Garc{\'{\i}}a}, {Masegosa},
  {Gonz{\'a}lez-Mart{\'{\i}}n}, {M{\'a}rquez}  \&
  {Perea}}{{Hern{\'a}ndez-Garc{\'{\i}}a} et~al.}{2016}]{2016Lore}
{Hern{\'a}ndez-Garc{\'{\i}}a} L.,  {Masegosa} J.,  {Gonz{\'a}lez-Mart{\'{\i}}n}
  O.,  {M{\'a}rquez} I.,   {Perea} J.,  2016, \mn@doi [\apj]
  {10.3847/0004-637X/824/1/7}, \href
  {http://adsabs.harvard.edu/abs/2016ApJ...824....7H} {824, 7}

\bibitem[\protect\citeauthoryear{{Ho}}{{Ho}}{2002}]{2002ho}
{Ho} L.~C.,  2002, \mn@doi [\apj] {10.1086/324399}, \href
  {https://ui.adsabs.harvard.edu/abs/2002ApJ...564..120H} {564, 120}

\bibitem[\protect\citeauthoryear{{Ho}}{{Ho}}{2008}]{2008ho}
{Ho} L.~C.,  2008, \mn@doi [\araa] {10.1146/annurev.astro.45.051806.110546},
  \href {http://adsabs.harvard.edu/abs/2008ARA%26A..46..475H} {46, 475}

\bibitem[\protect\citeauthoryear{{Ho}, {Filippenko}, {Sargent}  \& {Peng}}{{Ho}
  et~al.}{1997}]{1997Ho}
{Ho} L.~C.,  {Filippenko} A.~V.,  {Sargent} W.~L.~W.,   {Peng} C.~Y.,  1997, in
  American Astronomical Society Meeting Abstracts \#189. p.~735

\bibitem[\protect\citeauthoryear{{Ho}, {Greene}, {Filippenko}  \&
  {Sargent}}{{Ho} et~al.}{2009}]{2009bho}
{Ho} L.~C.,  {Greene} J.~E.,  {Filippenko} A.~V.,   {Sargent} W. L.~W.,  2009,
  \mn@doi [\apjs] {10.1088/0067-0049/183/1/1}, \href
  {https://ui.adsabs.harvard.edu/abs/2009ApJS..183....1H} {183, 1}

\bibitem[\protect\citeauthoryear{{Kalberla}, {Burton}, {Hartmann}, {Arnal},
  {Bajaja}, {Morras}  \& {Poeppel}}{{Kalberla} et~al.}{2005}]{2005karberla}
{Kalberla} P.~M.~W.,  {Burton} W.~B.,  {Hartmann} D.,  {Arnal} E.~M.,  {Bajaja}
  E.,  {Morras} R.,   {Poeppel} W.~G.~L.,  2005, VizieR Online Data Catalog,
  \href {http://adsabs.harvard.edu/abs/2005yCat.8076....0K} {8076}

\bibitem[\protect\citeauthoryear{{Keek} \& {Ballantyne}}{{Keek} \&
  {Ballantyne}}{2016}]{keek2016}
{Keek} L.,  {Ballantyne} D.~R.,  2016, \mn@doi [\mnras]
  {10.1093/mnras/stv2882}, \href
  {https://ui.adsabs.harvard.edu/abs/2016MNRAS.456.2722K} {456, 2722}

\bibitem[\protect\citeauthoryear{{Koratkar} \& {Blaes}}{{Koratkar} \&
  {Blaes}}{1999}]{1999Pkora}
{Koratkar} A.,  {Blaes} O.,  1999, \mn@doi [\pasp] {10.1086/316294}, \href
  {http://adsabs.harvard.edu/abs/1999PASP..111....1K} {111, 1}

\bibitem[\protect\citeauthoryear{{Krips} et~al.,}{{Krips}
  et~al.}{2005}]{2005krips}
{Krips} M.,  et~al., 2005, \mn@doi [\aap] {10.1051/0004-6361:20041731}, \href
  {https://ui.adsabs.harvard.edu/abs/2005A&A...442..479K} {442, 479}

\bibitem[\protect\citeauthoryear{{Krips} et~al.,}{{Krips}
  et~al.}{2007}]{2007krips}
{Krips} M.,  et~al., 2007, \mn@doi [\aap] {10.1051/0004-6361:20065037}, \href
  {https://ui.adsabs.harvard.edu/abs/2007A&A...464..553K} {464, 553}

\bibitem[\protect\citeauthoryear{{Li} \& {Xie}}{{Li} \& {Xie}}{2017}]{2017Li}
{Li} S.-L.,  {Xie} F.-G.,  2017, \mn@doi [\mnras] {10.1093/mnras/stx1778},
  \href {https://ui.adsabs.harvard.edu/abs/2017MNRAS.471.2848L} {471, 2848}

\bibitem[\protect\citeauthoryear{Lusso et~al.,}{Lusso et~al.}{2012}]{lusso2012}
Lusso E.,  et~al., 2012, \mn@doi [Monthly Notices of the Royal Astronomical
  Society] {10.1111/j.1365-2966.2012.21513.x}, 425, 623–640

\bibitem[\protect\citeauthoryear{{Ma}, {Yuan}  \& {Wang}}{{Ma}
  et~al.}{2007}]{2007ma}
{Ma} R.-Y.,  {Yuan} F.,   {Wang} D.-X.,  2007, \mn@doi [\apj] {10.1086/522917},
  \href {https://ui.adsabs.harvard.edu/abs/2007ApJ...671.1981M} {671, 1981}

\bibitem[\protect\citeauthoryear{{Maoz}}{{Maoz}}{2007}]{2007maoz}
{Maoz} D.,  2007, \mn@doi [\mnras] {10.1111/j.1365-2966.2007.11735.x}, \href
  {https://ui.adsabs.harvard.edu/abs/2007MNRAS.377.1696M} {377, 1696}

\bibitem[\protect\citeauthoryear{{Markakis} et~al.,}{{Markakis}
  et~al.}{2015}]{2015markakis}
{Markakis} K.,  et~al., 2015, \mn@doi [\aap] {10.1051/0004-6361/201425077},
  \href {https://ui.adsabs.harvard.edu/abs/2015A&A...580A..11M} {580, A11}

\bibitem[\protect\citeauthoryear{{McHardy}, {Koerding}, {Knigge}, {Uttley}  \&
  {Fender}}{{McHardy} et~al.}{2006}]{2006mchardy}
{McHardy} I.~M.,  {Koerding} E.,  {Knigge} C.,  {Uttley} P.,   {Fender} R.~P.,
  2006, \mn@doi [\nat] {10.1038/nature05389}, \href
  {http://adsabs.harvard.edu/abs/2006Natur.444..730M} {444, 730}

\bibitem[\protect\citeauthoryear{{Moustakas} \& {Kennicutt}}{{Moustakas} \&
  {Kennicutt}}{2006}]{2006Moustakas}
{Moustakas} J.,  {Kennicutt} Robert~C. J.,  2006, \mn@doi [\apjs]
  {10.1086/500971}, \href
  {https://ui.adsabs.harvard.edu/abs/2006ApJS..164...81M} {164, 81}

\bibitem[\protect\citeauthoryear{{Murphy} \& {Yaqoob}}{{Murphy} \&
  {Yaqoob}}{2009}]{2009Murphy}
{Murphy} K.~D.,  {Yaqoob} T.,  2009, \mn@doi [\mnras]
  {10.1111/j.1365-2966.2009.15025.x}, \href
  {http://adsabs.harvard.edu/abs/2009MNRAS.397.1549M} {397, 1549}

\bibitem[\protect\citeauthoryear{{Nagar}, {Falcke}  \& {Wilson}}{{Nagar}
  et~al.}{2005}]{2005neil}
{Nagar} N.~M.,  {Falcke} H.,   {Wilson} A.~S.,  2005, \mn@doi [\aap]
  {10.1051/0004-6361:20042277}, \href
  {https://ui.adsabs.harvard.edu/abs/2005A&A...435..521N} {435, 521}

\bibitem[\protect\citeauthoryear{{Nandra} \& {Pounds}}{{Nandra} \&
  {Pounds}}{1994a}]{Nandra1994}
{Nandra} K.,  {Pounds} K.~A.,  1994a, \mn@doi [\mnras]
  {10.1093/mnras/268.2.405}, \href
  {https://ui.adsabs.harvard.edu/abs/1994MNRAS.268..405N} {268, 405}

\bibitem[\protect\citeauthoryear{{Nandra} \& {Pounds}}{{Nandra} \&
  {Pounds}}{1994b}]{1994Nandra}
{Nandra} K.,  {Pounds} K.~A.,  1994b, \mn@doi [\mnras]
  {10.1093/mnras/268.2.405}, \href
  {https://ui.adsabs.harvard.edu/abs/1994MNRAS.268..405N} {268, 405}

\bibitem[\protect\citeauthoryear{{Nandra}, {George}, {Mushotzky}, {Turner}  \&
  {Yaqoob}}{{Nandra} et~al.}{1997}]{Nandra1997}
{Nandra} K.,  {George} I.~M.,  {Mushotzky} R.~F.,  {Turner} T.~J.,   {Yaqoob}
  T.,  1997, \mn@doi [\apjl] {10.1086/310937}, \href
  {https://ui.adsabs.harvard.edu/abs/1997ApJ...488L..91N} {488, L91}

\bibitem[\protect\citeauthoryear{{Nandra}, {O'Neill}, {George}  \&
  {Reeves}}{{Nandra} et~al.}{2007}]{2007nandra}
{Nandra} K.,  {O'Neill} P.~M.,  {George} I.~M.,   {Reeves} J.~N.,  2007,
  \mn@doi [\mnras] {10.1111/j.1365-2966.2007.12331.x}, \href
  {https://ui.adsabs.harvard.edu/abs/2007MNRAS.382..194N} {382, 194}

\bibitem[\protect\citeauthoryear{{Narayan}, {Yi}  \& {Mahadevan}}{{Narayan}
  et~al.}{1994}]{1994narayan}
{Narayan} R.,  {Yi} I.,   {Mahadevan} R.,  1994, arXiv Astrophysics e-prints,
  \href {http://adsabs.harvard.edu/abs/1994astro.ph.11060N} {}

\bibitem[\protect\citeauthoryear{{Netzer}}{{Netzer}}{1990}]{1990netzer}
{Netzer} H.,  1990, in {Blandford} R.~D.,  {Netzer} H.,  {Woltjer} L.,
  {Courvoisier} T.~J.~L.,   {Mayor} M.,  eds, Active Galactic Nuclei. pp
  57--160

\bibitem[\protect\citeauthoryear{{Netzer}}{{Netzer}}{2015}]{2015Netzer}
{Netzer} H.,  2015, \mn@doi [\araa] {10.1146/annurev-astro-082214-122302},
  \href {http://adsabs.harvard.edu/abs/2015ARA%26A..53..365N} {53, 365}

\bibitem[\protect\citeauthoryear{{Osorio-Clavijo}, {Gonz{\'a}lez-Mart{\'\i}n},
  {Papadakis}, {Masegosa}  \& {Hern{\'a}ndez-Garc{\'\i}a}}{{Osorio-Clavijo}
  et~al.}{2019}]{2019natalia}
{Osorio-Clavijo} N.,  {Gonz{\'a}lez-Mart{\'\i}n} O.,  {Papadakis} I.,
  {Masegosa} J.,   {Hern{\'a}ndez-Garc{\'\i}a} L.,  2019, arXiv e-prints, \href
  {https://ui.adsabs.harvard.edu/abs/2019arXiv191001660O} {p. arXiv:1910.01660}

\bibitem[\protect\citeauthoryear{{Osterbrock}}{{Osterbrock}}{1989}]{1989osterbrock}
{Osterbrock} D.~E.,  1989, {Astrophysics of gaseous nebulae and active galactic
  nuclei}

\bibitem[\protect\citeauthoryear{{Panessa}, {Barcons}, {Bassani}, {Cappi},
  {Carrera}, {Ho}  \& {Pellegrini}}{{Panessa} et~al.}{2007}]{2007panessa}
{Panessa} F.,  {Barcons} X.,  {Bassani} L.,  {Cappi} M.,  {Carrera} F.~J.,
  {Ho} L.~C.,   {Pellegrini} S.,  2007, \mn@doi [\aap]
  {10.1051/0004-6361:20066943}, \href
  {https://ui.adsabs.harvard.edu/abs/2007A&A...467..519P} {467, 519}

\bibitem[\protect\citeauthoryear{{Petrucci}, {Merloni}, {Fabian}, {Haardt}  \&
  {Gallo}}{{Petrucci} et~al.}{2001}]{petrucci2001}
{Petrucci} P.~O.,  {Merloni} A.,  {Fabian} A.,  {Haardt} F.,   {Gallo} E.,
  2001, \mn@doi [\mnras] {10.1046/j.1365-8711.2001.04897.x}, \href
  {https://ui.adsabs.harvard.edu/abs/2001MNRAS.328..501P} {328, 501}

\bibitem[\protect\citeauthoryear{{Piro}, {Yamauchi}  \& {Matsuoka}}{{Piro}
  et~al.}{1990}]{piro1990}
{Piro} L.,  {Yamauchi} M.,   {Matsuoka} M.,  1990, \mn@doi [\apjl]
  {10.1086/185806}, \href
  {https://ui.adsabs.harvard.edu/abs/1990ApJ...360L..35P} {360, L35}

\bibitem[\protect\citeauthoryear{{Pounds}, {Nandra}, {Stewart}, {George}  \&
  {Fabian}}{{Pounds} et~al.}{1990}]{1990Pounds}
{Pounds} K.~A.,  {Nandra} K.,  {Stewart} G.~C.,  {George} I.~M.,   {Fabian}
  A.~C.,  1990, \mn@doi [\nat] {10.1038/344132a0}, \href
  {https://ui.adsabs.harvard.edu/abs/1990Natur.344..132P} {344, 132}

\bibitem[\protect\citeauthoryear{{Rees}}{{Rees}}{1984}]{1984Rees}
{Rees} M.~J.,  1984, \mn@doi [\araa] {10.1146/annurev.aa.22.090184.002351},
  \href {http://adsabs.harvard.edu/abs/1984ARA%26A..22..471R} {22, 471}

\bibitem[\protect\citeauthoryear{Ricci, Tazaki, Ueda, Paltani, Boissay  \&
  Terashima}{Ricci et~al.}{2014}]{ricci2014suzaku}
Ricci C.,  Tazaki F.,  Ueda Y.,  Paltani S.,  Boissay R.,   Terashima Y.,
  2014, The Astrophysical Journal, 795, 147

\bibitem[\protect\citeauthoryear{{Ricci} et~al.,}{{Ricci}
  et~al.}{2017a}]{2017ricciapJS}
{Ricci} C.,  et~al., 2017a, \mn@doi [\apjs] {10.3847/1538-4365/aa96ad}, \href
  {https://ui.adsabs.harvard.edu/abs/2017ApJS..233...17R} {233, 17}

\bibitem[\protect\citeauthoryear{{Ricci} et~al.,}{{Ricci}
  et~al.}{2017b}]{2017Riccii}
{Ricci} C.,  et~al., 2017b, \mn@doi [\nat] {10.1038/nature23906}, \href
  {http://adsabs.harvard.edu/abs/2017Natur.549..488R} {549, 488}

\bibitem[\protect\citeauthoryear{{Ricci} et~al.,}{{Ricci}
  et~al.}{2018}]{2018Ricci}
{Ricci} C.,  et~al., 2018, \mn@doi [\mnras] {10.1093/mnras/sty1879}, \href
  {http://adsabs.harvard.edu/abs/2018MNRAS.480.1819R} {480, 1819}

\bibitem[\protect\citeauthoryear{{Risaliti}}{{Risaliti}}{2002}]{Risaliti2002}
{Risaliti} G.,  2002, \mn@doi [\aap] {10.1051/0004-6361:20020170}, \href
  {https://ui.adsabs.harvard.edu/abs/2002A&A...386..379R} {386, 379}

\bibitem[\protect\citeauthoryear{{Satyapal}, {Dudik}, {O'Halloran}  \&
  {Gliozzi}}{{Satyapal} et~al.}{2005}]{Satyapal2005}
{Satyapal} S.,  {Dudik} R.~P.,  {O'Halloran} B.,   {Gliozzi} M.,  2005, \mn@doi
  [\apj] {10.1086/449304}, \href
  {https://ui.adsabs.harvard.edu/abs/2005ApJ...633...86S} {633, 86}

\bibitem[\protect\citeauthoryear{{Shakura} \& {Sunyaev}}{{Shakura} \&
  {Sunyaev}}{1973}]{1973shakura}
{Shakura} N.~I.,  {Sunyaev} R.~A.,  1973, in {Bradt} H.,  {Giacconi} R.,  eds,
  IAU Symposium Vol. 55, X- and Gamma-Ray Astronomy. p.~155

\bibitem[\protect\citeauthoryear{{She}, {Ho}, {Feng}  \& {Cui}}{{She}
  et~al.}{2018}]{2018She}
{She} R.,  {Ho} L.~C.,  {Feng} H.,   {Cui} C.,  2018, \mn@doi [\apj]
  {10.3847/1538-4357/aabfe7}, \href
  {http://adsabs.harvard.edu/abs/2018ApJ...859..152S} {859, 152}

\bibitem[\protect\citeauthoryear{{Sparke}, {van Moorsel}, {Schwarz}  \&
  {Vogelaar}}{{Sparke} et~al.}{2009}]{2009sparke}
{Sparke} L.~S.,  {van Moorsel} G.,  {Schwarz} U.~J.,   {Vogelaar} M.,  2009,
  \mn@doi [\aj] {10.1088/0004-6256/137/4/3976}, \href
  {https://ui.adsabs.harvard.edu/abs/2009AJ....137.3976S} {137, 3976}

\bibitem[\protect\citeauthoryear{{Str{\"u}der} et~al.,}{{Str{\"u}der}
  et~al.}{2001}]{2001struder}
{Str{\"u}der} L.,  et~al., 2001, \mn@doi [\aap] {10.1051/0004-6361:20000066},
  \href {https://ui.adsabs.harvard.edu/abs/2001A&A...365L..18S} {365, L18}

\bibitem[\protect\citeauthoryear{{Terashima} \& {Wilson}}{{Terashima} \&
  {Wilson}}{2003}]{2003terashima}
{Terashima} Y.,  {Wilson} A.~S.,  2003, \mn@doi [\apj] {10.1086/345339}, \href
  {https://ui.adsabs.harvard.edu/abs/2003ApJ...583..145T} {583, 145}

\bibitem[\protect\citeauthoryear{{Titarchuk}}{{Titarchuk}}{1994}]{titarchuk1994}
{Titarchuk} L.,  1994, \mn@doi [\apj] {10.1086/174760}, \href
  {https://ui.adsabs.harvard.edu/abs/1994ApJ...434..570T} {434, 570}

\bibitem[\protect\citeauthoryear{{Tremaine} et~al.,}{{Tremaine}
  et~al.}{2002}]{tremaine2002}
{Tremaine} S.,  et~al., 2002, \mn@doi [\apj] {10.1086/341002}, \href
  {https://ui.adsabs.harvard.edu/abs/2002ApJ...574..740T} {574, 740}

\bibitem[\protect\citeauthoryear{{Tueller} et~al.,}{{Tueller}
  et~al.}{2010}]{tueller2010}
{Tueller} J.,  et~al., 2010, \mn@doi [\apjs] {10.1088/0067-0049/186/2/378},
  \href {https://ui.adsabs.harvard.edu/abs/2010ApJS..186..378T} {186, 378}

\bibitem[\protect\citeauthoryear{{Turner} et~al.,}{{Turner}
  et~al.}{2001}]{2001turner}
{Turner} M.~J.~L.,  et~al., 2001, \mn@doi [\aap] {10.1051/0004-6361:20000087},
  \href {https://ui.adsabs.harvard.edu/abs/2001A&A...365L..27T} {365, L27}

\bibitem[\protect\citeauthoryear{{Ueda} et~al.,}{{Ueda}
  et~al.}{2010}]{2010ueda}
{Ueda} Y.,  et~al., 2010, \mn@doi [\apj] {10.1088/0004-637X/713/1/257}, \href
  {https://ui.adsabs.harvard.edu/abs/2010ApJ...713..257U} {713, 257}

\bibitem[\protect\citeauthoryear{{Vaughan}, {Edelson}, {Warwick}  \&
  {Uttley}}{{Vaughan} et~al.}{2003}]{2003Vaughan}
{Vaughan} S.,  {Edelson} R.,  {Warwick} R.~S.,   {Uttley} P.,  2003, \mn@doi
  [\mnras] {10.1046/j.1365-2966.2003.07042.x}, \href
  {http://adsabs.harvard.edu/abs/2003MNRAS.345.1271V} {345, 1271}

\bibitem[\protect\citeauthoryear{{Xu}}{{Xu}}{2011}]{2011xu}
{Xu} Y.-D.,  2011, \mn@doi [\apj] {10.1088/0004-637X/729/1/10}, \href
  {https://ui.adsabs.harvard.edu/abs/2011ApJ...729...10X} {729, 10}

\bibitem[\protect\citeauthoryear{{Yamaoka}, {Uzawa}, {Arai}, {Yamazaki}  \&
  {Yoshida}}{{Yamaoka} et~al.}{2005}]{yamaoka2005}
{Yamaoka} K.,  {Uzawa} M.,  {Arai} M.,  {Yamazaki} T.,   {Yoshida} A.,  2005,
  Chinese Journal of Astronomy and Astrophysics Supplement, \href
  {https://ui.adsabs.harvard.edu/abs/2005ChJAS...5..273Y} {5, 273}

\bibitem[\protect\citeauthoryear{{Younes}, {Porquet}, {Sabra}  \&
  {Reeves}}{{Younes} et~al.}{2011}]{2011younes}
{Younes} G.,  {Porquet} D.,  {Sabra} B.,   {Reeves} J.~N.,  2011, \mn@doi
  [\aap] {10.1051/0004-6361/201116806}, \href
  {http://adsabs.harvard.edu/abs/2011A%26A...530A.149Y} {530, A149}

\bibitem[\protect\citeauthoryear{{Younes}, {Porquet}, {Sabra}, {Reeves}  \&
  {Grosso}}{{Younes} et~al.}{2012}]{2012younes}
{Younes} G.,  {Porquet} D.,  {Sabra} B.,  {Reeves} J.~N.,   {Grosso} N.,  2012,
  \mn@doi [\aap] {10.1051/0004-6361/201118299}, \href
  {https://ui.adsabs.harvard.edu/abs/2012A&A...539A.104Y} {539, A104}

\bibitem[\protect\citeauthoryear{{Younes}, {Ptak}, {Ho}, {Xie}, {Terasima},
  {Yuan}, {Huppenkothen}  \& {Yukita}}{{Younes} et~al.}{2019}]{2019younes}
{Younes} G.,  {Ptak} A.,  {Ho} L.~C.,  {Xie} F.-G.,  {Terasima} Y.,  {Yuan} F.,
   {Huppenkothen} D.,   {Yukita} M.,  2019, \mn@doi [\apj]
  {10.3847/1538-4357/aaf38b}, \href
  {https://ui.adsabs.harvard.edu/abs/2019ApJ...870...73Y} {870, 73}

\bibitem[\protect\citeauthoryear{{Young}, {McHardy}, {Emmanoulopoulos}  \&
  {Connolly}}{{Young} et~al.}{2018}]{2018young}
{Young} A.~J.,  {McHardy} I.,  {Emmanoulopoulos} D.,   {Connolly} S.,  2018,
  \mn@doi [\mnras] {10.1093/mnras/sty509}, \href
  {http://adsabs.harvard.edu/abs/2018MNRAS.476.5698Y} {476, 5698}

\bibitem[\protect\citeauthoryear{{Yuan}, {Taam}, {Misra}, {Wu}  \&
  {Xue}}{{Yuan} et~al.}{2007}]{yuan2007}
{Yuan} F.,  {Taam} R.~E.,  {Misra} R.,  {Wu} X.-B.,   {Xue} Y.,  2007, \mn@doi
  [\apj] {10.1086/511301}, \href
  {https://ui.adsabs.harvard.edu/abs/2007ApJ...658..282Y} {658, 282}

\makeatother
\end{thebibliography}

    \label{lastpage}

    \end{document}